\documentclass[aps,prx,reprint,superscriptaddress,longbibliography]{revtex4-1}
\usepackage{mathrsfs}
\usepackage{epsfig}
\usepackage{graphicx}
\usepackage{amsfonts}
\usepackage[figuresright]{rotating}
\usepackage{amssymb}
\usepackage{amsmath}
\usepackage{dcolumn}
\usepackage{bm}
\usepackage{color}
\usepackage[colorlinks, citecolor=blue]{hyperref}
\usepackage{verbatim}
\usepackage{amsmath,amssymb,amsfonts,bm}
\usepackage{amsthm,amsmath,amssymb}
\usepackage{MnSymbol}
\newcommand{\RNum}[1]{\uppercase\expandafter{\romannumeral #1\relax}}
\hypersetup{linkcolor=magenta,urlcolor=blue,citecolor=blue,pdfstartview={FitH},urlcolor=blue}

\def\be{\begin{equation}} \def\ee{\end{equation}}
\def\bea{\begin{eqnarray}} \def\eea{\end{eqnarray}}

\def\sign{\text{sign}}
\def\Tr{\text{Tr}}
\def\tr{\text{tr}}
\def\gtd{\tilde{g}}

\newcommand{\ket}[1]{| #1 \rangle}
\newcommand{\bra}[1]{\langle #1 |}
\newcommand{\Z}{\mathbb{Z}} 

\begin{document}

\title{Intrinsic Mixed-state Topological Order}

\author{Zijian Wang}
\thanks{These authors contributed equally to this work.}
\affiliation{Institute for Advanced Study, Tsinghua University, Beijing 100084,
People's Republic of China}
\author{Zhengzhi Wu}
\thanks{These authors contributed equally to this work.}
\affiliation{Institute for Advanced Study, Tsinghua University, Beijing 100084,
People's Republic of China}
\affiliation{Rudolf Peierls Centre for Theoretical Physics, Parks Road, Oxford, OX1 3PU, UK}
\author{Zhong Wang}
\email{wangzhongemail@tsinghua.edu.cn}
\affiliation{Institute for Advanced Study, Tsinghua University, Beijing 100084,
People's Republic of China}

\begin{abstract}  
Decoherence is a major obstacle to the preparation of topological order in noisy intermediate-scale quantum devices. Here, we show that decoherence can also give rise to new types of topological order. Specifically, we construct concrete examples by proliferating fermionic anyons in the toric code via local quantum channels. The resulting mixed states retain long-range entanglement, which manifests in the nonzero topological entanglement negativity, though the topological quantum memory is destroyed by decoherence. By comparison to the gapless spin liquid in pure states, we show that the identified states represent a novel intrinsic mixed-state topological order, which has no counterpart in pure states. Through the lens of quantum anomalies of 1-form symmetries, we then provide general constructions of intrinsic mixed-state topological order, and reveal the existence of non-bosonic deconfined anyons as another key feature of these novel phases. The extended meaning and characterization of deconfined excitations and their statistics in mixed states are clarified. Moreover, when these deconfined anyons have nontrivial braiding statistics, we prove that the mixed states cannot be prepared via finite-depth local quantum channels from any bipartite separable states. We further demonstrate our construction using the decohered Kitaev honeycomb model and the decohered double semion model. In the latter case, a surprising scenario arises where decoherence gives rise to additional types of deconfined anyons.      

\end{abstract}

\maketitle
\section{Introduction}
As long-range entangled (LRE) quantum matter, topologically ordered phases have attracted extensive attention in the past few decades \cite{wen1990topological,chen2010local,wen2017colloquium,savary2017spinliquid,zhou2017spinliquid,sachdev2018topological}. 
Recently, there is a growing number of theoretical proposals \cite{aguado2008creation,verresen2021prediction,piroli2021quantum,tantivasadakarn2021long,tantivasadakarn2022shortest,tantivasadakarn2023hierarchy,lee2022decoding,bravyi2022adaptive,lu2023mixed} as well as experimental evidence \cite{semeghini2021probing,satzinger2021realizing,google2023non,xu2023digital,iqbal2023topological,iqbal2023creation} showing that topological order (TO) can be prepared in current many-body quantum simulation platforms, such as superconducting qubit arrays, Rydberg atom arrays, and trapped ion systems, etc. A key feature of these noisy intermediate-scale quantum (NISQ) devices is the inevitable presence of decoherence, which renders the quantum state a mixed state \cite{preskill2018}. Substantial progress has been made in diagnosing nontrivial topological phases subject to decoherence \cite{dennis2002topological,Depolarization2012,mcginley2020fragility,deng2021stability,wang2021symmetry,de2022symmetry,lee2022symmetry,zhang2022strange,fan2023diagnostics,bao2023mixed,lee2023quantum,ma2023average,ma2023topological,bardyn2018probing,mao2023dissipation,su2023higher,wang2023topologically,liu2024dissipative}. Particularly, the decoherence-induced breakdown of topological quantum memory 
in the toric code model is investigated, and is related to the transition in the mixed-state topological order \cite{fan2023diagnostics,lee2023quantum}. 
The topological order therein is inherited from the pure-state counterpart, which is resistant to modest decoherence. Above certain critical error rate, the long-range entanglement is destroyed.  

We are interested in the following question: Other than destroying the pure-state topological order, can decoherence give rise to novel types of topological order that are intrinsically mixed? The possibility of such an intriguing scenario arises from new mechanisms of anyon proliferation provided by decoherence, distinct from anyon condensation in pure states \cite{Bais2009Condensate,Kong2014condensation,Burnell2018anyon}. As a starting point, we explore such possibility in the context of $\Z_2$ (toric-code) topological order, which comprises three types of anyon excitations, $e,m,$ and $f=e\times m$, \cite{anderson1973,read1989statistics,read1991largeN,wen1991mean,anderson1987RVB,senthil2000Z2,moessner2001dimer,kitaev2003fault}. In pure states, either $e$ or $m$ (both being self-bosons) can condense, leading to a topologically trivial Higgs/confined phase \cite{fradkin1979phase}. Correspondingly, proliferation of $e,m$ anyons induced by decoherence also destroys long-range entanglement. In contrast, $f$ anyons are self-fermions, and therefore they cannot condense in a pure state; instead, strong fluctuation of $f$ anyons typically leads to a gapless spin liquid, which remains LRE. This motivates us to study the fate of $\Z_2$ topological order when $f$ anyons proliferate under decoherence. 

Specifically, we study the behavior of the toric code model \cite{kitaev2003fault} under local quantum channels that solely create $f$ anyons. The topological quantum memory degrades to classical memory above certain decoherence threshold, aligning with previous studies. However, it turns out that the mixed state still possesses nontrivial quantum topological order even in the absence of quantum memory. To diagnose the mixed-state topological order, we employ the topological entanglement negativity (TEN) \cite{vidal2013TEN,castelnovo2013TEN,Ryu2016TEN,Ryu2016edge}, which is a natural generalization of the topological entanglement entropy (TEE) \cite{kitaev2006topological,levin2006detecting}. TEN has been utilized to probe topological order in thermal equilibrium \cite{hart2018entanglement, lu2020detecting,lu2022entanglement,lu2022characterizing} or under shallow-depth noise channels \cite{fan2023diagnostics}, effectively acting as a faithful indicator of topological quantum memory in those studied cases. However, we find that the TEN fails to reflect topological quantum memory in this scenario, and remains unchanged across the transition. Nevertheless, the nonzero TEN still points to the persistence of long-range entanglement, a hallmark of topological order. Moreover, the absence of topological quantum memory indicates that the identified topological order has no pure-state counterpart and, therefore, is termed ``intrinsic mixed-state topological order" here. 

We then provide further understanding of this peculiar result from the perspective of quantum anomalies. Crucially, the noisy channel proliferating $f$ anyons preserves an anomalous $1$-form symmetry generated by $f$ anyons \cite{Gaiotto2015generalized,wen2019emergent,mcgreevy2023generalized}. Through the anomalous $1$-form symmetry, we show that the intrinsic mixed-state TO in the decohered toric code supports deconfined fermionic anyons, with a detailed explanation of the extended meaning of deconfinement and fermionic statistics in mixed states. We generally prove that mixed states with anomalous $1$-form symmetries must be LRE. Moreover, we show that when the anomalous $1$-form symmetry is generated by anyons with nontrivial braiding statistics, any two complementary parts of the mixed state are LRE (for arbitrary bipartition), generalizing the conjecture ``mixed-state anomaly  $\Rightarrow$ multipartite non-separability" in \cite{lessa2024mixed}, which focuses on $0$-form symmetries. 
With this perspective, our construction can be  generalized to get other intrinsic mixed-state TO, characterized by deconfined anyons with nontrivial statistics.

After introducing the general construction, we give two more examples of intrinsic mixed-state TO, the decohered Kitaev honeycomb model and the decohered double semion model. In the latter example, we find more surprising features in mixed-state TO. In addition to peeling off anyons in the original TO, decoherence can even give birth to new types of deconfined anyons, which leads to a non-modular anyon theory in our case. 

The rest of the paper is organized as follows. In Section \ref{sec:decoheredTC} we construct an intrinsic mixed-state TO by proliferating $f$ anyons in the toric code, and reveal its properties through investigations of information quantities including the coherent information and the TEN. We also discuss similarities and disparities to the gapless spin liquid phase in pure states. In Section \ref{sec:generalities} we discuss general aspects of intrinsic mixed-state from the perspective of 1-form symmetry anomalies and deconfined excitations. In Section \ref{sec:generalizations} we generalize our construction of mixed-state quantum TO
to decohered Kitaev honeycomb model and decohered double semion model. We conclude with a discussion in Section \ref{sec:discussion}. We note that Secs. \ref{sec:generalities} and \ref{sec:decoheredDS} have been added after the initial version of our work appeared on arXiv. Meanwhile, Refs. \cite{sohal2024noisy,cheng2024towards} appeared, which study the mixed-state TO from the perspective of anomalous 1-form symmetries. The discussion regarding the nonmodular anyon theory in Sec. \ref{sec:decoheredDS} is inspired by these works.

\section{Decohered Toric code as intrinsic mixed-state TO\label{sec:decoheredTC}}
\subsection{The model}
\begin{figure}[htb]
 \centering
\includegraphics[width=0.9\linewidth]{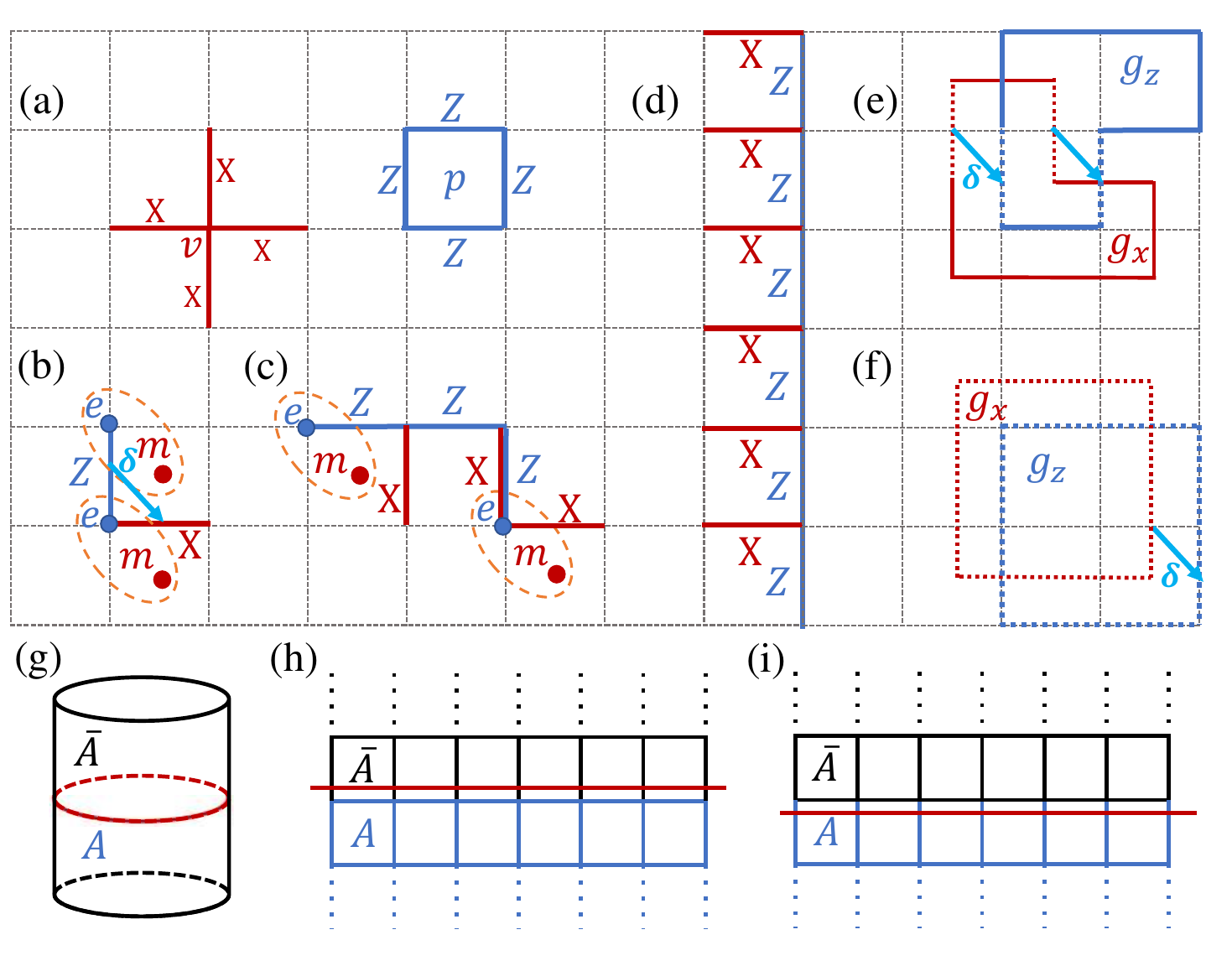}

\caption{(a). $A_v$ and $B_p$. (b). The Kraus operator $Z_iX_{i+\mathbf{\bm{\delta}}}$ of the quantum channel $\mathcal{N}^f_i$ creates a pair of $f$ anyons, labeled by orange ellipses. (c). An example of an error string $C$ (the blue line) and $W^f_C=\prod_{i\in C} Z_iX_{i+\mathbf{\bm{\delta}}}$. $f$ anyons are created at the ends of $C$. (d). A non-contractible loop operator $W_{\tilde{\gamma}_y}$ along $y$ direction. (e). A typical loop configuration $g=g_xg_z$, with $g_x$ colored in red and $g_z$ in blue. The dashed/solid lines represent segments where $g_x$, $g_z$ coincide/do not coincide (up to a shift by $\bm{\delta}$). $l_g=12$ in this example. (f). An example of a tensionless loop configuration, i.e., $l_g=0$. (g). Bipartition of a cylinder. (h)-(i) Two ways of bipartitions. }

\label{fig:fig1}
\end{figure}

We start with the 2D $\Z_2$ toric code model on a square lattice:
$$H_{\text{TC}}=-\sum_v A_v-\sum_p B_p,\ A_v\equiv\prod_{i\in v}X_i,\ B_p\equiv\prod_{i\in p}Z_i,$$ where $X_i,Z_i$ are Pauli matrices. The ground states are 4-fold degenerate and can be used to encode two logical qubits, amenable to fault-tolerant quantum information processing. The tolerance of the topological quantum memory against local phase errors and bit flip errors has been investigated in \cite{dennis2002topological,fan2023diagnostics,lee2023quantum}, where errors are modeled as local quantum channels,  $\mathcal{N}^z$ and $\mathcal{N}^x$,
\begin{equation}
\begin{aligned}
\mathcal{N}^{x}&=\prod_i\mathcal{N}_i^{x},\quad \mathcal{N}_i^{x}[\cdot]\equiv (1-p_{x})\cdot+p_{x}X_i\cdot X_i,\\
\mathcal{N}^{z}&=\prod_i\mathcal{N}_i^{z},\quad \mathcal{N}_i^{z}[\cdot]\equiv (1-p_{z})\cdot+p_{z}Z_i\cdot Z_i.
\end{aligned}
\end{equation}
$p_{x},p_z$ are the error rates of bit flip and phase errors respectively. Since these channels incoherently create $e$, $m$ anyons, respectively, we denote the corresponding error-corrupted states as $\rho_e$, $\rho_m$. It has been shown that above the error threshold, the proliferation of either bosonic anyon ($e$ or $m$) would degrade the quantum memory to classical memory, accompanied by a sudden drop of TEN from $\log 2$ to $0$. This motivates us to investigate the incoherent proliferation of the fermionic $f$ anyons of the $\Z_2$ topological order, which can be realized by the following 2-qubit quantum channel:

\begin{equation}
\mathcal{N}^{f}=\prod_i \mathcal{N}^{f}_i,\ \mathcal{N}^f_i[\cdot]:=(1-p_f)\cdot+p_f Z_iX_{i+\bm{\delta}}\cdot X_{i+\bm{\delta}}Z_i,
\label{eq:channel}
\end{equation}
where $\bm{\delta}=(\frac{1}{2},-\frac{1}{2})$ (the lattice constant is taken to be 1) and $0<p_f<\frac{1}{2}$ is the error rate. In this way the ground state $\rho_0$ is turned into a mixed state, $\rho_f=\mathcal{N}^f[\rho_0]$. As depicted in Fig.~\ref{fig:fig1}(b), $\mathcal{N}^f$ exclusively creates $f$ anyons. In contrast, certain other types of errors, like the Pauli-Y errors, locally create pairs of $f$ anyons as well, but globally they also produce $e$ and $m$ anyons, resulting in completely different outcomes. In the following sections, we demonstrate that this simple model surprisingly realizes an exotic intrinsic mixed-state topological order through analytical exact investigations of its topological memory and topological entanglement negativity.  

\subsection{Breakdown of quantum memory}
Under the $\mathcal{N}^f$ channel, the mixed state undergoes an error-induced transition corresponding to the breakdown of quantum memory, similar to the case with bit flip and phase errors. Such transitions can be probed by information quantities nonlinear in the density matrix, such as 
the coherent information $I_c=S(\rho_f)-S(\rho_{Rf})$ \cite{schumacher1996quantum,schumacher2001approximate,fan2023diagnostics}, where $S$ is the von Neumann entropy. $R$ denotes reference qubits purifying the initial state $\rho_{0}$, which is taken to be the maximally mixed state in the code space, 
\begin{equation}
\rho_0=\frac{1}{4}\prod_v\frac{1+A_v}{2}\prod_p\frac{1+B_p}{2}=\tr_R(|\Psi\rangle\langle\Psi|),
\label{eq:rho0}
\end{equation} 
and $\rho_{Rf}=\mathcal{I}_R\otimes \mathcal{N}^f[|\Psi\rangle\langle \Psi|]$ is the decohered density matrix. The coherent information measures the amount of information transmitted by the noisy channel $\mathcal{N}$,or in other words, it diagnoses the ability to restore the information encoded in the code space via error correction. Due to the subadditivity of the von Neumann entropy, the coherent information is bounded by $-S(\rho_0)\leq I_c\leq S(\rho_0)$, 
and the sufficient and necessary condition for the existence of perfect quantum error correction is that $I_c=S(\rho_{0})$ \cite{schumacher1996quantum}. The error threshold $p_c$ corresponding to a sudden drop of $I_c$ captures the breakdown of quantum memory, and this error rate threshold is an intrinsic threshold, which means for error rate $p>p_c$ there is no decoding algorithm to recover the encoded quantum information. 

In our model, we find that $I_c$ can be exactly mapped to the free energy cost of non-contractible defect lines of the random bond Ising model (RBIM) along the Nishimori line \cite{Nishimori1981internal}, using the replica trick:
\begin{equation}
\begin{aligned}
    I_c&=  -\lim_{n\rightarrow1}\frac{\partial }{\partial n}\Tr(\rho_{f}^n)+\lim_{n\rightarrow1}\frac{\partial }{\partial n}\Tr(\rho_{Rf}^n)\\
    &=2\log 2-\overline{\log \frac{\sum_{d_x,d_y=0,1 }Z^{\text{RBIM}}_{d_x,d_y}}{Z^{\text{RBIM}}_{00}}}\\
    &=2\log 2-\overline{\log\left[\sum_{d_x,d_y=0,1 }e^{-\Delta F_{d_x,d_y}}\right]},
\end{aligned}
\label{eq:Ic_mapping}
\end{equation}
where $\Delta F_{d_x,d_y}$ is the excess free energy with the insertion of a non-contractible defect line, and $d_x,d_y$ count the number of non-contractible defect lines in the $x$ and $y$ directions, respectively. The derivation of the above mapping of $I_c$  can be found in Appendix \ref{cohe_inf}. 

For small $p$, the RBIM is in the ferromagnetic (FM) phase, and the excess free energy of a defect line is extensive, $\Delta F_{\{d_x,d_y\}\neq\{0,0\}}\sim O(L)$, which leads to $I_c=2\log 2$. At a critical error rate  $p_c\approx 0.109$, the RBIM undergoes a ferromagnet-to-paramagnet phase transition, with an abrupt drop of coherent information, which determines the threshold where the topological quantum memory is damaged beyond recovery. Nevertheless, we note that $\rho_f$ still retains classical memory for $p_f>p_c$. Suppose the initial state $\rho_0$ is in an eigenspace of the logical operators $W_{\tilde{\gamma}_{x,y}}=\prod_{i\in \tilde{\gamma}_{x,y}}X_{i}Z_{i+\bm{\delta}}$, where $\tilde{\gamma}_{x},\tilde{\gamma}_{y}$ are two non-contractible loops on the dual lattice. Then $\rho_f$ always stays in the same eigenspace under the quantum channel, as $[W_{\tilde{\gamma}_{x,y}},Z_iX_{i+\bm{\delta}}]=0$.

\subsection{Topological entanglement negativity}
Based on the above analysis, it may seem that $\rho_f$ closely resembles $\rho_e$ and $\rho_m$. However, surprisingly, we demonstrate below that even when the quantum memory breaks down for $p_f > p_c$, $\rho_f$ retains LRE with a nonzero TEN, indicating the emergence of a distinct topological order.
 
To evaluate the entanglement negativity of $\rho_f$ and its scaling, we take the cylinder geometry with the bipartition $A\bigcup \bar A$ as depicted in Fig.~\ref{fig:fig1}(g). Then the logarithmic negativity is defined as:
\begin{equation}
\varepsilon_A(\rho_f) \equiv \log ||\rho_f^{T_A}||_1 = \varepsilon_{\bar A}(\rho_f),
\end{equation}
where $T_A$ denotes the partial transpose of $\rho_f$ in subregion A, and $||\cdot||_1$ represents the trace norm. As an entanglement monotone, the logarithmic negativity is commonly used to quantify quantum entanglement in mixed states, excluding the contribution from classical correlation \cite{separability,horodecki19961,zyczkowski1998volume,vidal2002computable}. As such, it is considered a natural generalization of entanglement entropy in pure states.

For convenience, we take the initial state $\rho_0$ to be the maximally mixed state in the code space \eqref{eq:rho0}.
We denote the groups generated by $\{A_v\}(\{B_p\})$ as $G_{x(z)}$:
\begin{equation}
G_x\equiv \langle \{A_v\}\rangle,\quad G_z\equiv\langle\{B_p\}\rangle.
\end{equation}
Each group element $g_{x(z)}$ corresponds to a loop configuration on the dual lattice (original lattice), as shown in Fig.~\ref{fig:fig1}(e), (f). Thus, $\rho_0$ can be represented by an equal weight expansion of loop configurations:
\begin{equation}
\begin{aligned}
    \rho_0=\frac{1}{2^N}\sum_{g_x\in G_x}\sum_{g_z\in G_z}g_xg_z=\frac{1}{2^N}\sum_{g\in G\equiv G_x\times G_z}g.
    \label{loop}
\end{aligned}
\end{equation}

The effect of $\mathcal{N}^f$ is to introduce loop tension. Specifically, for a given loop $g=g_xg_z$, $\mathcal{N}^f$ assigns a weight $1-2p_f$ to each segment where $g_x$ and $g_z$ does not coincide (up to a shift by $\bm{\delta}$). Consequently, we have,
\begin{equation}
\rho_f=\mathcal{N}^f[\rho_0]=\frac{1}{2^N}\sum_{g\in G} (1-2p_f)^{l_g}g,
\end{equation}
where $l_g$ is the length of segments where $g_x$ and $g_z$ do not coincide. In Fig.~\ref{fig:fig1}(e) we illustrate how to count such segments, and in Fig.~\ref{fig:fig1}(f) we give an example of a tensionless loop.

Now we take the partial transpose for subregion $A$. We denote $g=g_A g_{\bar A}$, where $g_{A(\bar A)}$ is the restriction of operator $g$ to subregion $A(\bar A)$,
\begin{equation}
\begin{aligned}
\rho_f^{T_A}=\frac{1}{2^N}\sum_{g\in G}(1-2p_f)^{l_g}y_A(g)g,
\end{aligned}
\label{tension}
\end{equation}
and $y_A(g)=1(-1)$ when $g_{xA}$ and $g_{zA}$ commute (anti-commute). 


As shown in Fig.~\ref{fig:fig1}(g)-(i) there are two possible choices of translation-invariant entanglement cut, which lead to slightly different results on TEN. Remarkably, for the bipartition in Fig.~\ref{fig:fig1}(h), the final result of negativity is rather simple, and is independent of $p_f$:
\begin{equation}
\varepsilon_A(\rho_f)=L\log 2-\log 2,
\end{equation}
where $L$ is the length of the boundary between $A$ and $\bar A$. For the bipartition in Fig.~\ref{fig:fig1}(i), however, the calculation of negativity is much harder for general $p_f$. Here we only show the results for the case with maximal decoherence, $p_f=\frac{1}{2}$, which is expected to reflect general features of the mixed states for $p_f>p_c$. In this case $\rho_f$ becomes the maximally mixed state with $W_p\equiv A_{p-\bm{\delta}}B_p=1,\forall p$. 
\begin{equation}
\rho_f=\frac{1}{2^{N/2+1}}\prod_p\frac{1+W_p}{2}
\label{eq:rhof_max1}
\end{equation}
It turns out the negativity exhibits an unusual dependence on the parity of $L$:
\begin{equation}
\varepsilon_A(\rho_f)=
\left\{
\begin{aligned}
\frac{L}{2}\log 2-\log2,\quad  & \text{if } L \text{ is even,} \\ 
\frac{L}{2}\log 2-\frac{\log2}{2},\quad  & \text{if } L \text{ is odd.}
\end{aligned}
\right. 
\label{eq:TEN_rough}
\end{equation}

In all the above results, the entanglement negativity satisfies an area law, and has an $O(1)$ subleading term, 
known as the topological entanglement negativity, which is a generalization of TEE. A nonzero value of TEN signals nontrivial quantum TO, as it arises solely from long-range entanglement. For example, TEN$=\log 2$ for the toric-code ground state. Hence, it has been used to diagnose topological order in both finite-temperature systems \cite{lu2020detecting} and states subject to local errors \cite{fan2023diagnostics}. Notably, in our model, the TEN remains nonzero even if quantum memory is gone, which is in sharp contrast to the case with single-qubit $X$ or $Z$ errors. 

The dependence of TEN on the boundary size for the second type of bipartition seems a bit puzzling. However, this phenomenon also shows up in some ground-state topological order. Namely, the TEE/TEN also exhibits similar even/odd dependence on system size for $\Z_2$ topological order enriched by translation symmetry through weak symmetry breaking, with typical examples including the Wen-plaquette model and the Abelian phase of Kitaev honeycomb model \cite{wen2003quantum,kitaev2006anyons}. In these cases, it is well known that the ground state degeneracy on a torus also exhibits similar dependence on the system size. Although there is little discussion in the literature about how weak symmetry breaking affects entanglement properties, it is straightforward to check that the subleading term of bipartite entanglement entropy on a cylinder also depends on the parity of the boundary size in these models. Based on this observation, we establish a connection between the entanglement properties of $\rho_f$ and ground-state $\Z_2$ topological order enriched by translation symmetry in Appendix \ref{weak_sym}. 

We emphasize that the persistence of long-range entanglement signifies genuine quantum TO, which distinguishes $\rho_{f}$ from the so-called classical TO, a concept raised in the study of finite-temperature TO \cite{castelnovo2007classical,castelnovo2007finiteT,castelnovo2008topological,lu2020detecting}. States with classical TO have topological classical memory as well, but zero TEN. One typical example is the low-temperature phase of the 3D toric code model. 
In this sense, $\rho_e$ and $\rho_m$ (above the error threshold) also have classical TO. $\rho_f$ is qualitatively different from these known examples. 

 \subsection{Robustness of the intrinsic mixed-state TO}\label{sec:robustness} 
\begin{figure}[htb]
 \centering
\includegraphics[width=0.8\linewidth]{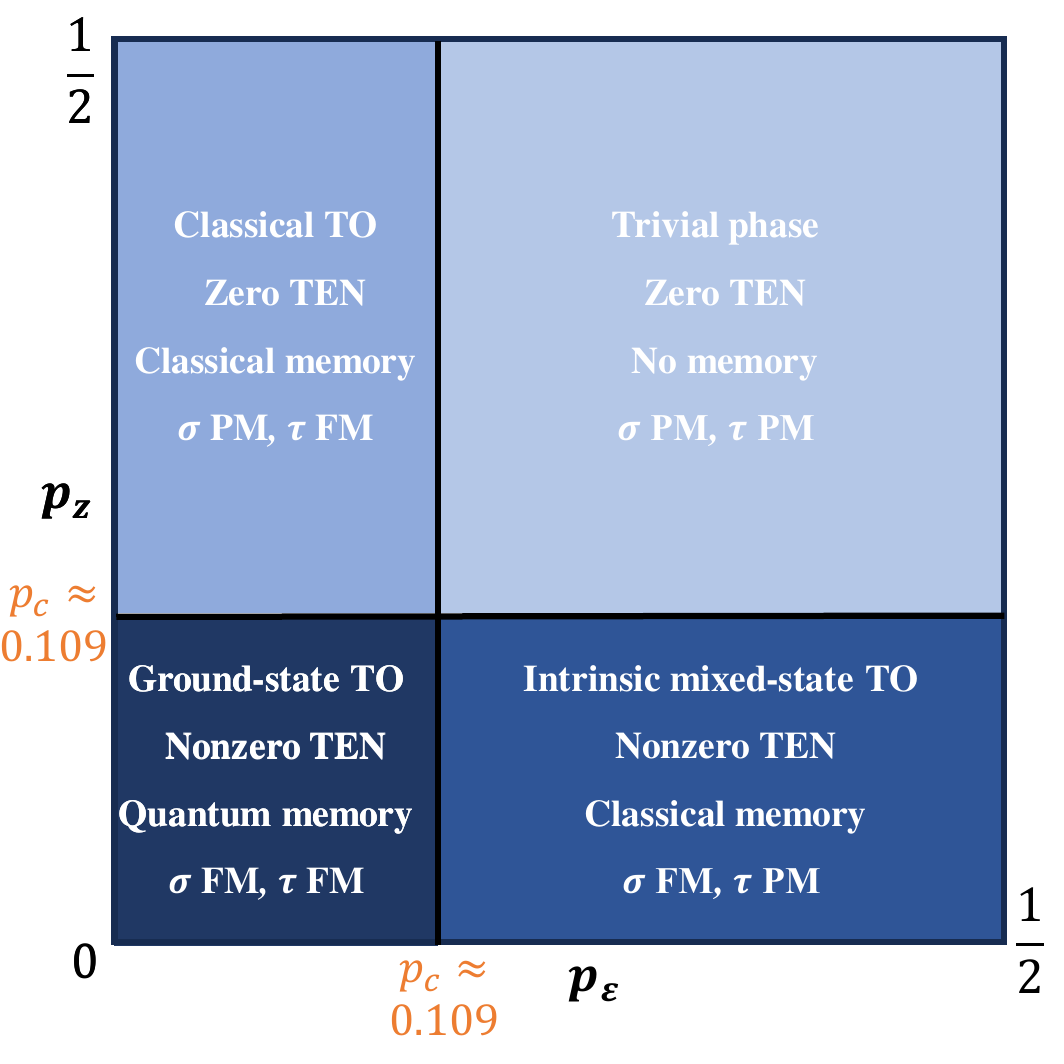}

\caption{The phase diagram of the toric code model subjected to both two-qubit errors and single-qubit phase errors. Properties of each phase (including topological memory and TEN) and the corresponding phases in the two RBIMs (with Ising variables $\sigma$,$\tau$, respectively) are indicated in the phase diagram.}

\label{fig:fig3}
\end{figure}

 Although in our construction we need to use specific $2$-qubit channels that look a bit unconventional, $\rho_{f}$ represents a new type of topologically ordered phase, instead of a fine-tuned exception. We can consider the case when single-qubit phase errors are also present: ${\rho}_{f,e}=\mathcal{N}^z[\rho_f]$, with error rate $p_z$. By mapping to two decoupled RBIMs, we obtain the phase diagram in Fig.~\ref{fig:fig3}. For small $p_z$, ${\rho}_{f,e}$ stays in the same phase as $\rho_f$, while for $p_z>p_c\approx 0.109$, the state undergoes another transition to the trivial phase, with no memory and zero TEN. See Appendix \ref{rob_app} for details.

\subsection{Comparison to anyon condensation and gapless spin liquid}
From the previous analysis, we see that although $\rho_f$ only exhibits classical memory (for $p_f>p_c$), it is fundamentally distinct from $\rho_e$ and $\rho_m$. To gain deeper insight into this counter-intuitive result, we draw comparisons between the error-induced anyon proliferation and anyon condensation in pure states. Instead of applying local quantum channels, we analyze the case when the $X_i,Z_i,Z_iX_{i+\bm{\delta}}$ terms are directly introduced into the toric code Hamiltonian: 
\begin{equation}
H=H_{\text{TC}}-\sum_ih_xX_i-h_zZ_i-h_{xz}Z_iX_{i+\bm{\delta}},
\end{equation} 

The ground state phase diagram for $h_{xz}=0$ has been extensively studied \cite{fradkin1979phase,trebst2007breakdown,Tupitsyn2010topological,dusuel2011robustness,Nahum2021selfdual}. For sufficiently large $h_z$ or $h_x$, it leads to the condensation of $e$ or $m$ anyons, respectively, resulting in the destruction of long-range entanglement. Analogously, the local $Z,X$ errors induce $e,m$ anyon proliferation in an incoherent manner, which also destroys the long-range entanglement. Despite the similarities between decoherence-induced anyon proliferation and anyon condensation, there are still noteworthy distinctions. In mixed states, incoherent proliferation of either $e$ or $m$ does not completely trivialize the phase but rather leads to classical TO. Contrarily, in pure states, condensation of either $e$ or $m$ already leads to the trivial Higgs/confined phase. 

The distinction becomes much more significant for $f$ anyons. As fermions, they cannot condense in pure states. Then what happens when we turn on fluctuations of $f$ anyons ($h_{zx}\neq 0$)? Surprisingly, at $h_z=h_x=0$, this model can be exactly solved via fermionization \cite{chen2018bosonization}. As we show in Appendix \ref{app:fermionization}, when $h_{xz}$ is sufficiently large ($h_{x,z}=0$), corresponding to large fluctuations of $f$ particles, the system enters a gapless spin liquid phase, where $f$ particles form a $p-$wave superconductor with conic dispersion, similar to the gapless Kitaev spin liquid \cite{kitaev2006anyons}. The intrinsic mixed-state TO proposed in this paper has many common features with the gapless spin liquid. First, they are both obtained by proliferating $f$ anyons in the $\Z_2$ TO, and, as a result, they both have no quantum memory. For the gapless spin liquid, it is due to the absence of a spectrum gap, so there is no well-defined topological degeneracy. Second, despite the lack of quantum memory, they are both LRE, and are thus nontrivial phases of matter. In the next section we will uncover a deeper reason of their LRE nature from the perspective of anomalies. 
However, there are also notable differences between $\rho_{f}$ and the gapless spin liquid phase. Gapless phases typically exhibit critical behaviors, including algebraically decaying correlation functions and subleading logarithmic corner contributions to entanglement entropy/negativity \cite{corner2006,corner2009,corner2015}. Contrarily, since $\rho_{f}$ is obtained by applying local quantum channels on a gapped topological order, no power-law correlation can be generated, i.e., $\rho_f$ exhibits short-range correlation for all local operators, which is a prerequisite for any topological order. Moreover, as we have seen, the subleading term in the entanglement negativity $\varepsilon_A(\rho_f)$ is always $O(1)$ for $\rho_{f}$. In this sense, $\rho_{f}$ also retains certain essential properties of gapped topological order. Therefore, $\rho_{f}$ indeed represents a new type of topological order that is only possible in mixed states. The similarities and differences between gapless spin liquid and intrinsic mixed-state TO are summarized in Table \ref{Table:1}.
\begin{table}[htb]
\begin{tabular}{|p{2.5cm}<{\centering}|p{1.9cm}<{\centering}|p{1.9cm}<{\centering}|p{1.9cm}<{\centering}|}
\hline
 & Gapless spin liquid & Intrinsic mixed-state TO \\ \hline
Quantum memory & $\times$ &  $\times$   \\\hline
 Long-range entanglement & $\surd$ & $\surd$    \\ \hline
 Correlation of local operators & Power-law correlation & Short-range correlation\\ \hline
 Anomalous 1-form symmetry & $\surd$ & $\surd$     \\ \hline
 Deconfined fermions & $\surd$ & $\surd$  \\\hline
\end{tabular}
\caption{Comparison between the gapless spin liquid and the intrinsic mixed-state TO ($\rho_f$). See Section \ref{sec:generalities} for properties listed in the last two rows} \label{Table:1}
\end{table}

\section{ Generalities: anomalous 1-form symmetry, nontrivial statistics, and long-range entanglement\label{sec:generalities}}

The long-range entanglement of the gapless spin liquid and the intrinsic mixed-state TO are both related to an anomalous 1-form symmetry, which is generated by the following loop operators
\begin{equation}
W^f_{\tilde{\gamma}}=\prod_{i\in \tilde{\gamma}}X_iZ_{i+\mathbf{\delta}}, 
\end{equation}
where $\tilde{\gamma}$ denotes an arbitrary loop on the dual lattice \cite{Gaiotto2015generalized,wen2019emergent}. For non-contractible loops $\tilde{\gamma}=\tilde{\gamma}_{x,y}$, $W^f_{\tilde{\gamma}}$ are the logical operators responsible for the classical memory of $\rho_f$ (see \ref{fig:fig1}(d)). These loop operators generate a symmetry of both models because $[W^f_{\tilde{\gamma}},H_{\text{TC}}]=[W^f_{\tilde{\gamma}},Z_iX_{i+\bm{\delta}}]=0$. Particularly, we have $W^f_{\tilde{\gamma}} \rho_f=\rho_f$ \footnote{This equality always holds for contractible $\tilde{\gamma}$, while for non-contractible $
\tilde{\gamma}$ it only holds for a particular choice of the initial state $\rho_0$: $W_{\tilde{\gamma}}\rho_0=\rho_0$. Fortunately, for the argument below we only need to use this condition for contractible $\tilde{\gamma}$, so this subtlety can be safely ignored.} , which is known as the strong symmetry condition for the mixed state \cite{Buca2012symmetry,Lieu2020symmetry,de2022symmetry,ma2023average}. For an open string $\tilde{C}$, $W^f_{\tilde{C}}=\prod_{i\in \tilde{C}}X_iZ_{i+\bm{\delta}}$ creates two $f$ anyons at the ends of the string, so they are referred to as $f$ strings, and the $1$-form symmetry is said to be generated by $f$ anyons. The nontrivial statistics of $f$ anyons indicates an anomaly of the 1-form symmetry \cite{wen2019emergent}. We show below that, as a consequence of the anomalous strong 1-form symmetry, the mixed state $\rho_{f}$ still has deconfined fermionic excitations. 

The above statement might be confusing at first sight. Since the $f$ anyons already proliferate in $\rho_{f}$, what are the deconfined fermions then? Perhaps the easiest way to resolve this apparent paradox is to vectorize $\rho$ in the double Hilbert space: $\rho=\sum_{mn}\rho_{mn}|m\rangle\langle n|\rightarrow |\rho\rrangle=\sum_{mn}\rho_{mn}|m\rangle_+\otimes |n\rangle_{-}$. For concreteness we choose the basis $\{|m\rangle\}$ as eigenstates of $Z_i$. Then $|\rho_0\rrangle$ corresponds to two copies of toric-code ground states, with superselection sectors $\{1,e_+,m_+,f_+\}\times \{1,e_-,m_-,f_-\}$. The incoherent proliferation of $f$ corresponds to condensation of $f_+f_-$ in this picture, which leaves $f_+$ a deconfined excitation, though it should be now identified with $f_-$ due to the condensation of $f_+f_-$ \cite{bao2023mixed}. Do these $f_+$ anyons correspond to physical excitation in the original Hilbert space? Although the naive way to create $f_+$ anyons $\rho_f\rightarrow W^f_{\tilde{C}}\rho$  is not a legitimate physical process, they can be created using the unitary process $\rho_f\rightarrow \rho'_f=U_{\tilde{C}}\rho_f U^\dagger_{\tilde{C}}$, with $U_{\tilde{C}}=\frac{I+iW^f_{\tilde{C}}}{\sqrt{2}}$. Then $f_{+/-}$ anyons appear in the interference terms. In this sense, $f$ anyons (we omit the ``$+$'' hereafter) remain physically meaningful excitations. 

Now we address another subtle question: What does ``deconfinement" really mean in mixed states under noisy channels? Conventionally, it means that the energy cost does not grow indefinitely by separating individual topological excitations far apart. Here this definition does not make sense because the system is no longer governed by a Hamiltonian. Therefore, we propose the following definition of deconfined excitations for generic mixed states.

\textbf{Definition 1.} Given a density matrix $\rho$, a pair of deconfined excitations are said to be created at locations $i,j$ in the unitary process $\rho\rightarrow U\rho U^\dagger$, iff the following two conditions are satisfied:
\begin{enumerate}
\item The excitations cannot be created locally and individually. In other words, $U$ cannot be any unitary operator supported near $i,j$. Typically, $U$ is supported on an open string with endpoints $i,j$. 
\item For any local operator $O$ whose support is away from $i,j$, $\text{tr}(\rho O)=\text{tr}(U\rho U^\dagger O)$. It means that the change can only be detected near $i,j$. 
\end{enumerate}

Clearly, the above definition is consistent with the conventional notion of deconfined excitations, and serves as a faithful and natural generalization to open quantum systems. Next, we illustrate that $f$ anyons (we omit the ``+'' hereafter) are indeed deconfined excitations according to this definition, created by the unitary operator $U_{\tilde{C}}$. The second condition can be easily verified using the strong $1$-form symmetry. We denote the support of $O$ as $\Omega(O)$ for convenience, which we assume to be away from $i,j$. If $\Omega(O)\bigcap \tilde{C}=\emptyset$, then $\tr(U_{\tilde{C}}\rho U_{\tilde{C}}^\dagger O)=\tr(\rho U_{\tilde{C}}^\dagger O U_{\tilde{C}})=\tr(\rho O)$; if $\Omega(O)\bigcap \tilde{C}\neq \emptyset$, we can always find another open string $\tilde{C}'$, such that $\tilde{C}\bigcup\tilde{C}'$ is a contractible loop and $\Omega(O)\bigcap\tilde{C}'=\emptyset$. Using the strong $1$-form symmetry $U_{\tilde{C}'}^\dagger U_{\tilde{C}}\rho_f=\rho_f$, it is straightforward to get $\tr(U_{\tilde{C}}\rho U_{\tilde{C}}^\dagger O)=\tr(U_{\tilde{C}'}\rho U_{\tilde{C}'}^\dagger O)=\tr(\rho O)$. The first condition follows from the fermionic statistics of the $f$ anyons \cite{levin2003fermions}. See also the proof of Theorem 1 for more details.

\begin{figure}[htb]
 \centering
\includegraphics[width=1.0\linewidth]{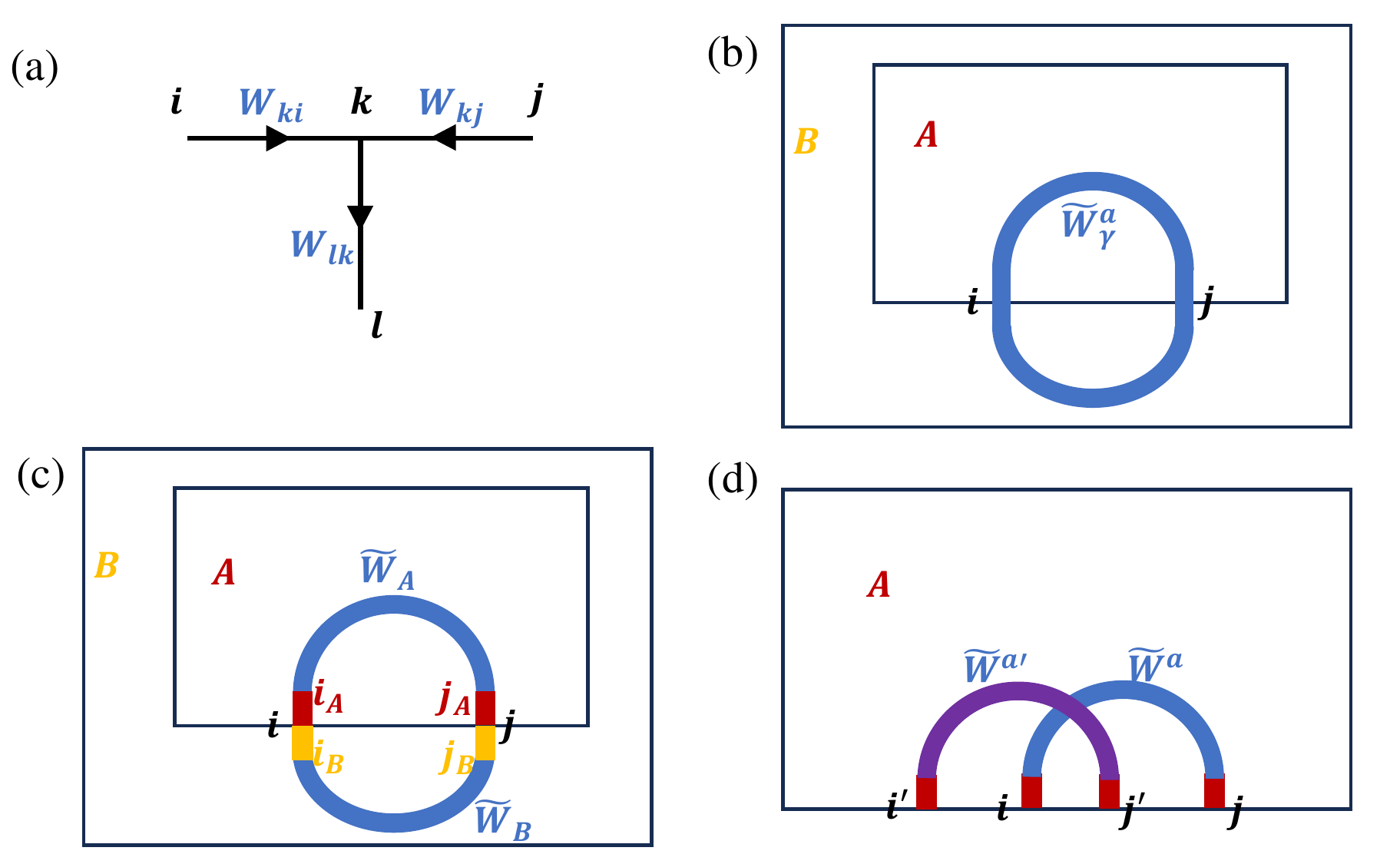}

\caption{(a). The exchange statistics of identical Abelian anyons can be defined using the open anyon strings $W_{ki},W_{kj},W_{lk}$. In defining open anyon strings, we require that two short strings can combine into a longer string: $W_{i_3i_2}W_{i_2i_1}=W_{i_3i_1}$. (b). A closed $a$ string that intersects $\partial A$ at $i,j$. (c). Partition of $W^a_\gamma$ into three parts: the open string $W_{i_Aj_A}$, $W_{j_Bi_B}$ supported completely on region $A,B$, and the middle red and orange segments $W_{AB}$ straddling between $A$ and $B$. (d). Anyon strings of $a$ and $a$'. }

\label{fig:exchange}
\end{figure}
Is the fermionic statistics well-defined for mixed states? We give an affirmative answer with the following microscopic method of detection (see Fig.~\ref{fig:exchange} for an illustration): $a.$ Create two $f$ anyons at locations $i$, $j$ using $W^f_{ji}=W^f_{jk}W^f_{ki}$; $b.$ move the $f$ anyon at $i$ to location $l$ using $W^f_{li}=W^f_{lk}W^f_{ki}$. $c.$ move the other $f$ anyon from $j$ to $i$ using $W^f_{ij}=W^{f\dagger}_{ji}$; $d.$ move the $f$ anyon at $l$ to location $j$ using $W^f_{jl}=W^f_{jk}W^f_{kl}$; $e.$ annihilate the two $f$ anyons using $W^f_{ij}$. 
In step $b$-$d$ the locations of the two $f$ anyons are exchanged, which result in a phase $W^f_{jl}W^f_{ij}W^f_{li}=\theta(f)=-1$. To turn this statistical phase into an observable effect, we can use the same protocol as in \cite{satzinger2021realizing}. That is, we introduce an ancilla qubit and prepare the initial state $|+\rangle\langle +|\otimes \rho_f$, where $|+\rangle\equiv\frac{|1\rangle+|0\rangle}{\sqrt{2}}$. Then we can use the ancilla qubit to control the exchange process. Namely, we perform step $a$-$e$ when the ancilla is in state $|1\rangle$, and do nothing otherwise. This controlled process can be performed using the unitary operator $V=|0\rangle\langle0|\otimes I+|1\rangle\langle 1|\otimes W^f_{ij}W^f_{jl}W^f_{ij}W^f_{li}W^f_{ji}$. Then the statistical phase will manifest as the rotation of the ancilla:
\begin{equation}
V\left( |+\rangle\langle +|\otimes\rho_f\right)V^\dagger=|-\rangle\langle-|\otimes\rho_f,
\label{eq:exchange}
\end{equation}
where $|-\rangle=\frac{|0\rangle-|1\rangle}{\sqrt{2}}$. The fermionic statistics of emergent deconfined excitations is the most striking observable effect of the intrinsic mixed-state TO. Here, it is guaranteed by the anomalous strong $1$-form symmetry. 

In the double space, $e_+e_-$ also remains a deconfined excitation, which has mutual semion statistics with $f_+$. Surprisingly, this braiding statistics can also be (partially) detected in the physical Hilbert space, as we demonstrate below. 
First, note that $\rho_f$ preserves a weak $1$-form symmetry generated by $e$ anyons, $W^e_\gamma\rho_f W^e_\gamma=\rho$ for contractible closed $e$ string $W^e_\gamma$ \cite{de2022symmetry}. Then based on Definition 1, deconfined excitations can be created using an open $e$ string, $\rho_f\rightarrow \rho^e_f=W^e_C\rho_fW^e_C$. Secondly, introduce an ancilla qubit and prepare the initial state $|+\rangle\langle +|\otimes \rho_f^{e}$. Then use the ancilla qubit to control the braiding process: if the ancilla is in $|1\rangle$, we create two $f$ anyons and drag one of them along a loop $l$ enclosing the $e$ anyon, and finally annihilate with the other $f$ anyon; if the ancilla is in $|0\rangle$ we do nothing. This controlled process can be performed using the unitary gate $V=|0\rangle\langle0|\otimes I+|1\rangle\langle 1|\otimes W^f_{\tilde{\gamma}}$, where $W^f_{\tilde{\gamma}}=\prod_{i\in \text{loop }\tilde{\gamma}}X_iZ_{i+\bm{\delta}}$. Then the braiding statistics can be detected by rotation of the ancilla qubit:
 \begin{equation}
 V\left( |+\rangle\langle +|\otimes\rho_f^{e}\right)V^\dagger=|-\rangle\langle-|\otimes\rho_f^{e}.
 \end{equation}

Stated more formally, the braiding statistics reflects the mixed anomaly between the weak $1$-form symmetry (generated by $e$) and the strong $1$-form symmetry (generated by $f$) \cite{wang2024anomaly}. 

One key distinction to the usual anyon braiding is that the $f-e$ braiding here is only one-way defined under the above protocol. Namely, the statistical phase can only be detected by moving $f$ around $e$ (with the movement of $f$ controlled by the ancilla), but not the other way around \footnote{Nevertheless, one can alternatively use the ancilla to control the creation of two $f$ anyons and extract the braiding statistics by moving $e$ around $f$.}. This directly follows from the fact that while $f$ anyons generate a strong $1$-form symmetry, $e$ anyons only generate a weak $1$-form symmetry. Therefore, a finer characterization of deconfined excitations than Definition 1 is needed, and we call the former type as strongly deconfined and the latter type as weakly deconfined. As demonstrated above, strongly deconfined anyons are allowed to form coherent superposition states, and their statistics can be detected from the interference effects. Weakly deconfined anyons, on the other hand, always have trivial exchange and braiding statistics among themselves and can have one-way braiding statistics with strongly deconfined anyons. In the rest of the paper we mainly focus on strongly deconfined anyons.
 \begin{figure}[htb]
  \centering
\includegraphics[width=0.7\linewidth]{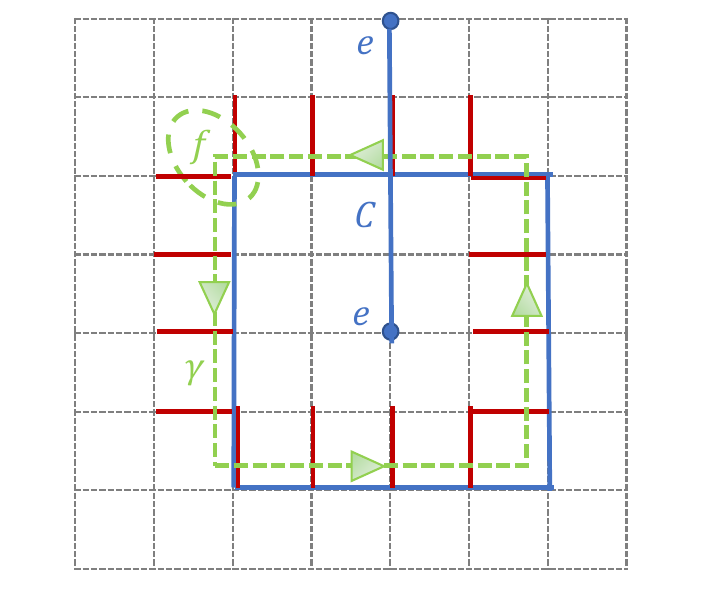}

 \caption{Braiding between $f$ and $e$. The quantum channel $\mathcal{N}^f$ does not affect the braiding statistics because it preserves the weak $1$-form symmetry generated by $e$ and the strong $1$-form symmetry generated by $f$.}

 \label{fig:braiding}
 \end{figure}

Using the perspective of anomalies, we can generalize our construction to obtain other intrinsic mixed-state TO \cite{sohal2024noisy,cheng2024towards}. Given a TO with anyon content $\mathcal{A}$, one can incoherently proliferate a subset $\mathcal{B}\subset\mathcal{A} $ of anyons using a noisy channel, taking the Kraus operator to be the shortest string of anyons in $\mathcal{B}$. If there exists some Abelian anyon $a\in \mathcal{A}$ that have trivial braiding with all anyons in $\mathcal{B}$,
then closed $a$ strings commute with the Kraus operators. Therefore, the $a$ anyons generate a strong $1$-form symmetry \cite{li2024replica}. Based on the previous discussion, it means they remain deconfined excitations with well-defined statistics. When they have nontrivial statistics, $\theta(a)\neq 1$, then the 1-form symmetry is anomalous. For example, in the model discussed above, $\mathcal{A}=\{1,e,m,f\},\mathcal{B}=\{f\}$, and $a=f$ because fermions have trivial full braiding with themselves. 

The existence of non-bosonic strongly deconfined excitation can be viewed as another diagnosis of nontrivial mixed-state TO, in complementary to TEN. Indeed, we show below that this property generically implies the mixed state to be LRE, without explicit calculation of negativity.

\textbf{Theorem 1:} Consider a $2$-dimensional state $\rho$ with anomalous strong 1-form symmetry generated by some Abelian anyon $a$ with $\theta(a)\neq 1$, i.e.,  $W^a_{\gamma}\rho=\rho$, where $W^a_\gamma$'s are closed $a$ strings supported on any contractible loop $\gamma$. Then $\rho$ cannot be prepared using a finite-depth local channel (FDLC) from any fully separable state:
\begin{equation}
\rho\neq \mathcal{N}_{\text{FDLC}}[\sum_{\lambda}p_\lambda\bigotimes_{\text{site }i} |\psi^\lambda_i\rangle \langle\psi^\lambda_i|]\ (p_\lambda>0).
\label{eq:ful_separable}
\end{equation}

Before the proof, we give some remarks regarding the above theorem.
\begin{enumerate}
\item Following \cite{hastings2011topological,sang2023mixed}, \eqref{eq:ful_separable} can be viewed as the definition of LRE for generic mixed states. Thus the above theorem tells us anomalous strong 1-form symmetries must lead to nontrivial mixed-state quatum TO. 
\item Typically, the 1-form symmetry includes generators supported on non-contractible loops. However, here we only require the symmetry condition for contractible ones, enabling a much wider range of applications. For example, for the construction of intrinsic mixed-state TO discussed above, the pre-decoherence state can be taken to be any state in the ground state subspace. The above theorem can also be applied to topologically trivial spatial manifolds like a $2$-sphere.  
\item Similar to the discussion above \eqref{eq:exchange}, the anomaly manifests as the algebra of open string operators: 
\begin{equation}
\begin{aligned}
&W^a_{i_3i_2}W^a_{i_2i_1}=W^a_{i_3i_1},W^\dagger_{i_1i_2}=W_{i_2i_1},\\
&W^a_{kj}W^a_{lk}W^a_{km}=\theta(a)W^a_{km}W^a_{lk}W^a_{kj} 
\end{aligned}
\label{eq:exchange_a}
\end{equation}
Essentially, the nontrivial statistics $\theta(a)\neq 1$ of $a$ is all we need in the following proof.
\item    Notably, the condition we impose automatically includes the case of mixed anomaly between strong 1-form symmetries, which arises when two strongly deconfined Abelian anyons $a$, $a'$ have nontrivial mutual statistics, $B_\theta(a,a')\neq 1$ ($B_\theta(a,a')$ is the statistical phase from a full braiding between $a$ and $a'$). Using the relation $B_\theta(a,a')=\frac{\theta(aa')}{\theta(a)\theta(a')}$, the mixed anomaly implies at least one of $a,a',aa'$ has nontrivial self statistics $\theta\neq 1$. Thus we do not need to consider mutual statistics separately. 
\item To prove the theorem, we make an additional assumption that $W^a_\gamma$ can be expressed using an FDLUC, which is usually the case for Abelian anyon $a$. 

\end{enumerate}

{\it Proof.} Suppose by contradiction that $\rho= \mathcal{N}_{\text{FDLC}}[\sum_{\lambda}p_\lambda\bigotimes_{\text{site }i} |\psi^\lambda_i\rangle \langle\psi^\lambda_i|]$. Recall that a generic FDLC $\mathcal{N}$ can be constructed in three steps: First, introduce auxiliary degrees of freedom on each site. Secondly, apply a finite-depth local unitary circuit (FDLUC) on the system and the auxiliary degrees of freedom. Finally, trace out the added degrees of freedom. It can be written as $\mathcal{N}_{\text{FDLC}}[\rho_0]=\tr_E[U_{SE}\rho_0\bigotimes_{\text{site i}}|e_i\rangle\langle e_i|U^\dagger_{SE}]$, where $U_{SE}$ is a FDLUC. If $\rho_0$ is a fully separable state, then $\rho=\mathcal{N}_{\text{FDLC}}[\rho_0]$ is of the form $\rho=\sum_\lambda p_\lambda \tr_E[U_{SE}\bigotimes_i|\phi_i^\lambda\rangle\langle \phi_i^\lambda|U^\dagger_{SE}]$, where $|\phi_i\rangle\equiv |\psi_i^\lambda\rangle\otimes|e_i^\lambda\rangle$ Then each pure state $U_{SE}\bigotimes_i|{\phi}_i\rangle$ must be symmetric: $W^a_\gamma U_{SE}\bigotimes_i|\phi_i\rangle=U_{SE}\bigotimes_i|\phi_i\rangle$. Equivalently, $\tilde{W}^a_\gamma=U^\dagger_{SE}{W}_\gamma^a U_{SE}$ is a symmetry of the product state $\bigotimes_i|\phi_i\rangle$. Crucially, the restriction of $\tilde{W}^a$ to open strings $\tilde{W}^a_{i_1i_2}\equiv U^\dagger_{SE}{W}_{i_1i_2}^a U_{SE}$ satisfies the same algebra \eqref{eq:exchange_a} as $W^a$. As we show below, this will lead to contradiction.

Since any contractible loop operator $\tilde{W}^a_\gamma$ is a symmetry, an open string $\tilde{W}^a_{i_1i_2}$ can only change the state near the end of the string: $\tilde{W}_{i_1i_2}\bigotimes_i|\phi_i\rangle=A_{i_1}B_{i_2}\bigotimes_i|\phi_i\rangle$, where $A_{i_1},B_{i_2}$ are unitary operators supported near $i_1,i_2$, respectively. From the algebraic relation \eqref{eq:exchange_a}, 
\begin{equation}
    \tilde{W}^a_{kj}\tilde{W}^a_{lm}=\theta(a)\tilde{W}^a_{km}\tilde{W}^a_{lj}.
\end{equation}
By applying both sides of the equation on $\bigotimes_i|\phi_i\rangle$, we get $A_k\otimes B_j\otimes A_l\otimes B_m\bigotimes_i|\phi_i\rangle=\theta(a)A_k\otimes B_m\otimes A_l\otimes B_j\bigotimes_i|\phi_i\rangle$, which leads to contradiction when $\theta(a)\neq 1$. $\qed$

The above proof shows that nontrivial TO is guaranteed by the existence of non-bosonic deconfined excitation $a$, even for mixed states. Furthermore, when $a$ is neither bosonic or fermionic, $\theta(a)\neq \pm 1,$ or when $a$ has nontrivial mutual statistics with another strongly deconfined anyon $a'$, then an even stronger conclusion can be proved.

\textbf{Theorem 2:} Consider a 2-dimensional state $\rho$ with an anomalous strong 1-form symmetry generated by Abelian anyons $a$, $a'$ with nontrivial braiding statistics $B_{\theta}(a,a')\neq 1$ (with the special case $a=a'$ included). Then for any bipartition $A \bigcup B$ with linear size $L_A, L_B \rightarrow \infty$ in the thermodynamic limit, $\rho$ cannot be prepared using a FDLC from any bipartite separable state: 

\begin{equation}
\rho\neq \mathcal{N}_{\text{FDLC}}[\sum_{\lambda}p_\lambda|\psi^\lambda_A\rangle \langle\psi^\lambda_A|\otimes |\psi^\lambda_B\rangle\langle \psi^\lambda_B|].
\label{eq:separable}
\end{equation}
Again, we only need the symmetry condition $W^a_\gamma\rho=\rho,W^{a'}_\gamma\rho=\rho$ for contractible loops operators $W^a_\gamma,W^{a'}_\gamma$, which we assume to be FDLUC's. 

{\it Proof.} Suppose, by contradiction, that $\rho= \mathcal{N}_{\text{FDLC}}[\sum_{\lambda}p_\lambda|\psi^\lambda_A\rangle \langle\psi^\lambda_A|\otimes |\psi^\lambda_B\rangle\langle \psi^\lambda_B|]$. Following the same steps as in the proof of Theorem 1, we have 
\begin{equation}
\tilde{W}^a_\gamma|\phi_A\rangle\otimes|\phi_B\rangle=|\phi_A\rangle\otimes|\phi_B\rangle, 
\label{eq:1-symmetry}
\end{equation}
with the bipartite product state $|\phi_A\rangle\otimes|\phi_B\rangle$ and the 1-form symmetry generator $\tilde{W}^a_\gamma$ defined in an enlarged Hilbert space $\mathcal{H}_S\otimes\mathcal{H}_E$. Next, we take a loop $\gamma$ intersecting the boundary between $A$ and $B$ at locations $i,j$, with the corresponding closed $a$ string $\tilde{W}^a_\gamma$. We extract a small segment of $\tilde{W}^a_\gamma$ near the intersection points, denoted by $\tilde{W}_{AB}$, such that $\tilde{W}^a_\gamma$ can be written as $\tilde{W}^a_\gamma=\tilde{W}^a_{i_Aj_A}\tilde{W}^a_{j_Bi_B}\tilde{W}_{AB}$, where $\tilde{W}^a_{i_Aj_A},\tilde{W}^a_{j_Bi_B}$ are open $a$ strings completely supported on $A,B$, respectively. Then we have 
\begin{equation}
\begin{aligned}
&\tilde{W}_{AB}|\phi_A\rangle\otimes|\phi_B\rangle=\tilde{W}^{a\dagger}_{i_Aj_A}|\phi_A\rangle \otimes \tilde{W}^{a\dagger}_{j_Bi_B}|\phi_B\rangle\\
\Rightarrow & \tr_{B\bigcup \sigma_{ij}}[ |\phi_A\rangle\langle \phi_A|\otimes|\phi_B\rangle\langle\phi_B|]
\\&=\tr_{B\bigcup \sigma_{ij}}[\tilde{W}^{a\dagger}_{i_Aj_A}|\phi_A\rangle\langle\phi_A| \tilde{W}^{a}_{i_Aj_A}\otimes |\phi_B\rangle\langle\phi_B|]\\
\Rightarrow &\tr_{\sigma_{ij}}[ |\phi_A\rangle\langle \phi_A|]=\tr_{\sigma_{ij}}[\tilde{W}^{a\dagger}_{i_Aj_A}|\phi_A\rangle\langle\phi_A|\tilde{W}^{a}_{i_Aj_A}],
\end{aligned}
\end{equation}
 where $\sigma_{ij}$ denotes the union of the two red segments in Fig.~\ref{fig:exchange}(c). Due to the unitary equivalence of purification \cite{watrous2018theory}, 
 \begin{equation}
 \begin{aligned}
 &\exists u_{\sigma_{ij}}\text{ supported on }\sigma_{ij}, \text{s.t. }u_{\sigma_{ij}}|\phi_A\rangle =\tilde{W}^{a\dagger}_{i_Aj_A}|\phi_A\rangle\\
 \Rightarrow
 & u^\dagger_{\sigma_{ij}}\tilde{W}^{a}_{j_Ai_A}|\phi_A\rangle=|\phi_A\rangle
 \end{aligned}
 \label{eq:eliminate1}
 \end{equation}
Physically, this means a pair of $a$ and its antiparticle are created in the bulk of $A$, moving apart towards $\partial A$, and get eliminated on the boundary (by a unitary process with local support). Below we show such process contradicts with the anyonic statistics of $a$ \cite{levin2012braiding,levin2013protected}.  
Now consider the closed $a'$ string $W^{a'}_{\gamma'}$ which intersects the boundary $\partial A$ at $i',j'$, with the distance between $i,j,i',j'$ sufficiently large compared to the depth of $\mathcal{N}_{\text{FDLC}}$. Similarly to the above analysis, 
 \begin{equation}
 \exists u_{\sigma_{i'j'}}\text{ supported on }\sigma_{i'j'}, \text{s.t. }u^\dagger_{\sigma_{i'j'}}\tilde{W}^{a}_{j'_Ai'_A}|\phi_A\rangle=|\phi_A\rangle.
 \label{eq:eliminate2}
 \end{equation}
 On the other hand, from the braiding statistics between $a,a'$:
 \begin{equation}
 \begin{aligned}
&\tilde{W}^{a'}_{j'_Ai'_A}\tilde{W}^a_{j_Ai_A}=B_\theta(a,a')\tilde{W}^a_{j_Ai_A}\tilde{W}^{a'}_{j'_Ai'_A}\\
\Rightarrow 
& u^\dagger_{\sigma_{i'j'}}\tilde{W}^{a'}_{j'_Ai'_A}u^\dagger_{\sigma_{ij}}\tilde{W}^{a}_{j_Ai_A}=B_\theta(a,a') u^\dagger_{\sigma_{ij}}\tilde{W}^{a}_{j_Ai_A}u^\dagger_{\sigma_{i'j'}}\tilde{W}^{a'}_{j'_Ai'_A},
\end{aligned}
 \end{equation}
which is inconsistent with \eqref{eq:eliminate1}, \eqref{eq:eliminate2} when $B_\theta(a,a')\neq 1$. $\qed$

We can restate Theorem 2 in the following way: if a mixed state has strongly deconfined Abelian anyons with nontrivial braiding statistics, then for any bipartition, there must be long-range entanglement between the two complementary regions. Notably, our discussion includes the scenario $a'=a$, in which case $B_\theta(a,a')=\theta^2(a)$. Compared to Theorem 1, we leave behind the case that $a$ has fermionic self-statistics but trivial mutual braiding statistics with other deconfined anyons. In such cases, $a$ also generates an anomalous strong $1$-form symmetry, but our proof does not work. Nevertheless, the decohered toric code under ``$ZX$" errors we construct is indeed bipartite LRE, indicated by the nonzero TEN, which arguably should also be related to the anomalous strong $1$-form symmetry generated by $f$ anyons. It is intruguing to explore whether the existence of deconfined fermions also generically lead to bipartite LRE mixed states.

The above results can be viewed as a generalization of the results in \cite{lessa2024mixed}, where it is conjectured that mixed states with anomalous strong $0$-form symmetry in $d$ spatial dimensions cannot be prepared via a FDLC from any $(d+2)$-partite separable states. In two dimensions it means the mixed states cannot be prepared via a FDLC from a $4$-partite non-separable state, with the additional condition that three of the four parts intersect at one point. Our results show that anomalies of $1$-form symmetries can have a stronger constraining power, since bipartite non-separability implies multipartite non-separability, but not vice versa.  



\section{Generalizations to other intrinsic mixed-state TO\label{sec:generalizations}} 
\subsection{Decohered Kitaev honeycomb model}
In the last section, we provide a general route to generalize the construction in Section \ref{sec:decoheredTC}  to obtain other intrinsic mixed-state TO. In this section, we give two more examples as applications. We first discuss the generalization to the Kitaev honeycomb model: $H=-J_{x} \sum_{x \text {-bonds }} \sigma_{j}^{x} \sigma_{k}^{x}-J_{y} \sum_{y \text {-bonds }} \sigma_{j}^{y} \sigma_{k}^{y}-J_{z} \sum_{z \text {-bonds }} \sigma_{j}^{z} \sigma_{k}^{z}-\sum_i\vec{h}\cdot\vec{\sigma}_i$ (with $|\vec{h}|\ll |J_\mu|, \mu=x,y,z$). This model can be exactly solved by mapping it to Majorana fermions coupled to static $\Z_2$ gauge fields \cite{kitaev2003fault}. It is shown that the ground state $\rho_0$ of this model can realize Abelian $\Z_2$ TO, non-Abelian Ising TO, as well as a gapless $\Z_2$ spin liquid phase. All three phases have deconfined fermion excitations.
\begin{figure}[htb]
 \centering
\includegraphics[width=0.8\linewidth]{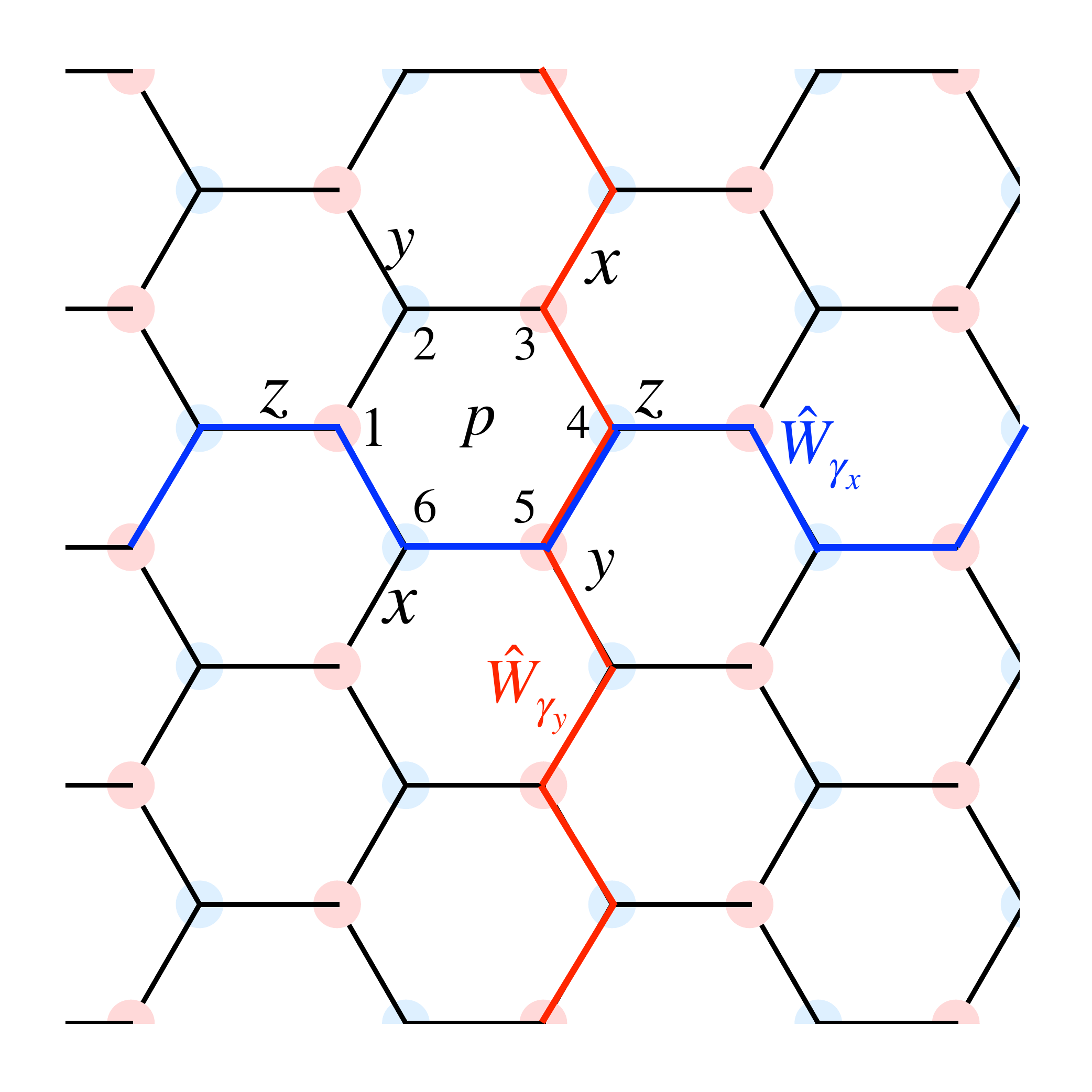}

\caption{The Kitaev honeycomb model, logical operators and flux operators. All the bonds are classified into three different equivalence classes of parallel bonds: x-bonds, y-bonds and z-bonds. There is a two-spin interaction $\sigma_i^{\alpha}\sigma_j^{\alpha}$ on each $\alpha$-bond, where $\alpha=x,y,z$. The two non-contractible loop operators $W_{\gamma_x}, W_{\gamma_y}$ illustrated as blue and red are the product of the two-spin interactions on the bonds along those loops. In each plaquette p, there is a conserved flux operator $W^f_p=\sigma_1^z\sigma_2^y\sigma_3^x\sigma_4^z\sigma_5^y\sigma_6^x$.   }

\label{fig:mto_kitaev}
\end{figure}

To obtain intrinsic mixed-state TO, we construct the following channel:
\begin{equation}
\begin{aligned}
&\rho_f=\mathcal{N}^X\circ\mathcal{N}^Y\circ\mathcal{N}^Z[\rho_0],\quad \mathcal{N}^\alpha=\prod_{\langle ij\rangle\in \alpha\text{-bonds}}\mathcal{N}^\alpha_{\langle ij\rangle},\\
 &\mathcal{N}^\alpha_{\langle ij\rangle}[\rho_0]= p\sigma^\alpha_i\sigma^\alpha_j\rho_0\sigma^\alpha_j\sigma^\alpha_i +(1-p)\rho_0.
\end{aligned}
\label{decoheredhoneycomb}
\end{equation}
This channel leads to decoherence of the fermions but preserves the $\Z_2$ gauge flux. In other words, it preserves the anomalous $1$-form symmetry, with generators $W^f_p=\prod_{\langle ij\rangle\in p}\sigma^z_1\sigma^y_2\sigma^x_3\sigma^z_4\sigma^y_5\sigma^x_6$ \cite{liu2024symmetries}. Therefore, the resulting mixed states must be LRE, and support deconfined fermionic excitations. In the maximally decohered case $p_f=\frac{1}{2}$, the $f$ particles are heated to infinite temperature, so all of the three ground-state phases will end up in the maximally mixed state in the zero-flux sector ($W^f_p=1,\forall p$), which belongs to the same intrinsic mixed-state TO as that constructed in Section \ref{sec:decoheredTC}. Actually, at $p_f=\frac{1}{2}$, this state can be obtained by applying a Hadamard gate on all vertical links to \eqref{eq:rhof_max1}. Therefore, the decohered honeycomb model constructed here is also characterized by a nonzero TEN. We note that a similar model in the context of Lindblad equations was constructed in \cite{hwang2023mixed}, but the LRE nature was not uncovered.  


\subsection{Decohered double semion model\label{sec:decoheredDS}}
As another example, we construct intrinsic mixed-state TO from the double semion TO. The anyon content of the double semion TO is $\mathcal{A}=\{1,s,\bar{s},s\bar{s}\}=\{1,s\}\times\{1,\bar s\}$, where $s$ is a semion, $\theta(s)=i$; $\bar{s}$ is an antisemion, $\theta(\bar{s})=-i$; and $s\bar{s}$ is a boson, $\theta(s\bar{s})=1$. The fusion rules are $s\times s=1,\bar{s}\times \bar{s}=1,s\times\bar{s}=s\bar{s}$ \cite{levin2005string}. Following the general strategy in Section \ref{sec:generalities}, we can proliferate the semion $s$ using noisy channels, and due to the trivial braiding between $s$ and $\bar{s}$, the $1$-form symmetry generated by $\bar{s}$ is preserved. Thus $\bar{s}$ remains a strongly deconfined excitation with well-defined anti-semionic statistics.
\begin{figure}[htb]
 \centering
\includegraphics[width=1.0\linewidth]{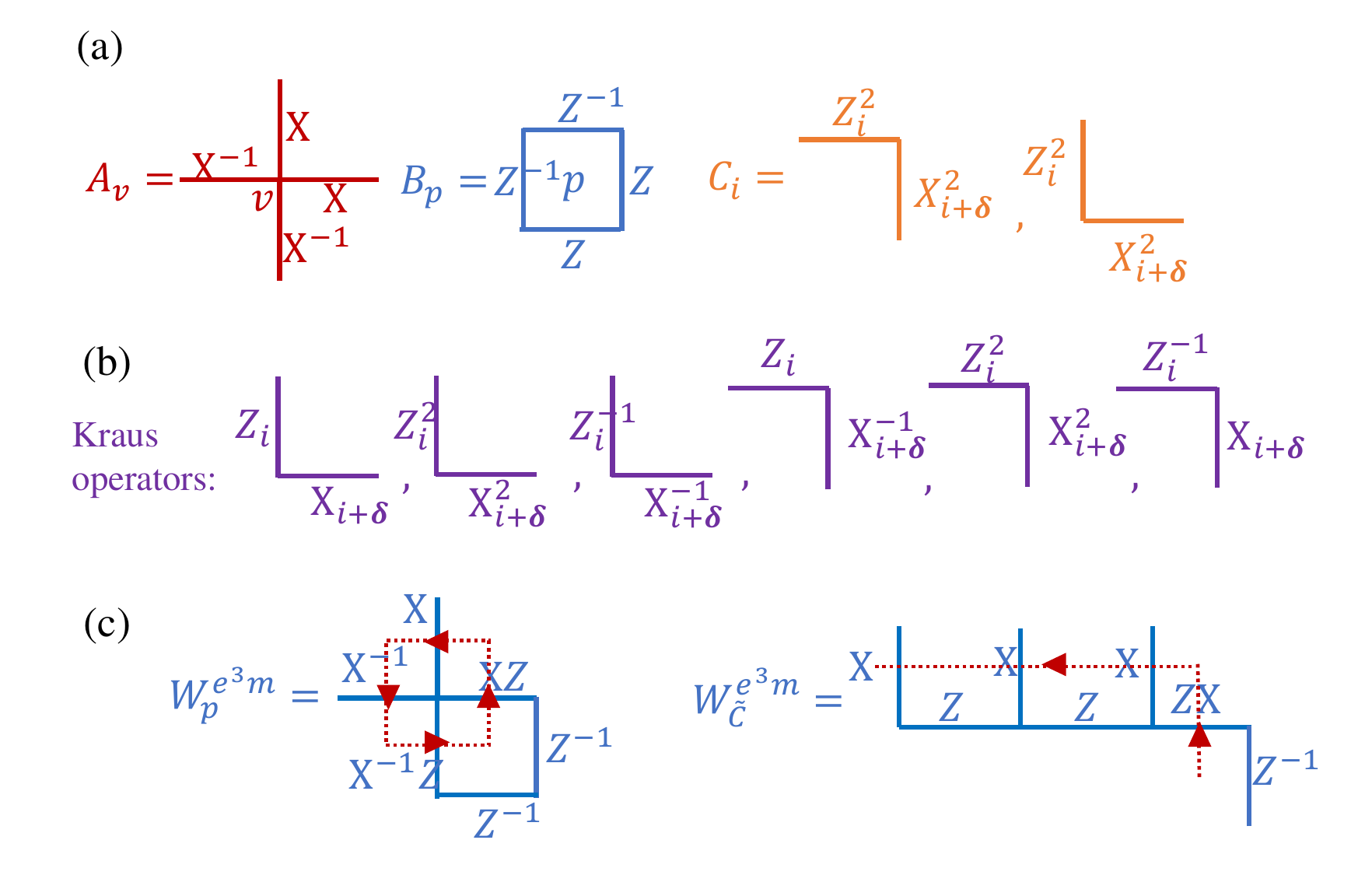}

\caption{(a). The stabilizers used in defining $H_{\Z_4 \text{ TC}}$ and  $H_{\text{DS}}$. (b). The Kraus operators in the decohered double semion model. They are the shortest string operators of $[em]$ anyons. (c). A shortest closed $e^3m$ string, $W_p^{e^3m}$, and an open $e^3m$ string $W^{e^3m}_{\tilde{C}}$.  }

\label{fig:DS}
\end{figure}

For concreteness, we start with the Pauli stabilizer model realizing the double semion TO \cite{Ellison2022twisted}. The model is defined on a 2D square lattice with a $\Z_4$ degree of freedom on each link, which is equipped with the Pauli operators $Z=\sum_{n\in \Z_4}i^n|n\rangle\langle n|$ and $X=\sum_{n\in\Z_4}|n+1\rangle\langle n|$. The stabilizer model is defined as follows:

\begin{equation}
H_{\text{DS}}=-\sum_p (A_{v=p-\bm{\delta}}B^{-1}_p+h.c.)-\sum_p B^2_p-\sum_i C_i,
\label{eq:H_DS}
\end{equation}
where $A_v,B_p,C_i$ are represented graphically in Fig.~\ref{fig:DS}(a). Below we briefly review how this model is constructed from a parent $\Z_4$ toric code:
\begin{equation}
H_{\Z_4 \text{ TC}}=-\sum_v A_v-\sum_p B_p+h.c.
\end{equation}
The anyon content of the $\Z_4$ toric code is $\{e^n m^r(n,r=0,1,2,3)\}$ with a $\Z_4\times \Z_4$ fusion rule, $e^4=m^4=1$. The statistics of the sixteen anyons are given by $\theta(e^nm^r)=i^{nr}$. The double semion TO described by $H_{\text{DS}}$ is obtained by condensing the boson $e^2m^2$ in the $\Z_4$ toric code (the $C_i$ term is the shortest string operator of $e^2m^2$, leading to its condensation), which causes confinement of anyons with nontrivial braiding with $e^2m^2$ and identification of anyons related by fusion with $e^2m^2$. Namely, $[a]=[a\times e^2m^2]$ in the condensed theory, where we use $[a]$ to label the remaining deconfined anyons after the $e^2m^2$ condensation. The deconfined anyons in the condensed theory are $\{[1],[em],[e^3m],[e^2]\}=\{[1],[em]\}\times\{[1],[e^3m]\}$, where $[em]$ is a semion and $[e^3m]$ is an anti-semion. Indeed, a double semion TO is realized.

Next, we investigate the effect of proliferating $[em]$ anyons using the following quantum channel,
\begin{equation}
\begin{aligned}
\mathcal{N}^{[em]}=\prod_i \mathcal{N}^{[em]}_i,\ \mathcal{N}^{[em]}_i[\cdot]\equiv\sum_{n=0,1,2,3}p_nK^n_i\cdot K^{n\dagger}_i.\\
\end{aligned}
\label{eq:emchannel}
\end{equation}
The Kraus operators $K_i$, $K^\dagger_i$ are the shortest string operators of $em$ anyons in the $Z_4$ toric code:
\begin{equation}
K_i=
\left\{
\begin{aligned}
&Z_iX_{i+\bm{\delta} } \text{   for vertical link } i \\
&Z_iX^{-1}_{i+\bm{\delta}}  \text{ for horizontal link } i
\end{aligned}
\right.
\end{equation}
For simplicity, we take the initial state $\rho_0$ to be the maximally mixed state in the ground-state subspace of ${H}_{\text{DS}}$ and directly consider the maximally decohered case $p_0=p_1=p_2=p_3=\frac{1}{4}$. Based on the analysis at the beginning of this section, it seems only the anti-semion $[e^3m]$ will remain strongly deconfined, which leads to a chiral anti-semion theory $\{1,[e^3m]\}$. However, the actual situation turns out to be even more intriguing. 

\begin{figure}[htb]
 \centering
\includegraphics[width=1\linewidth]{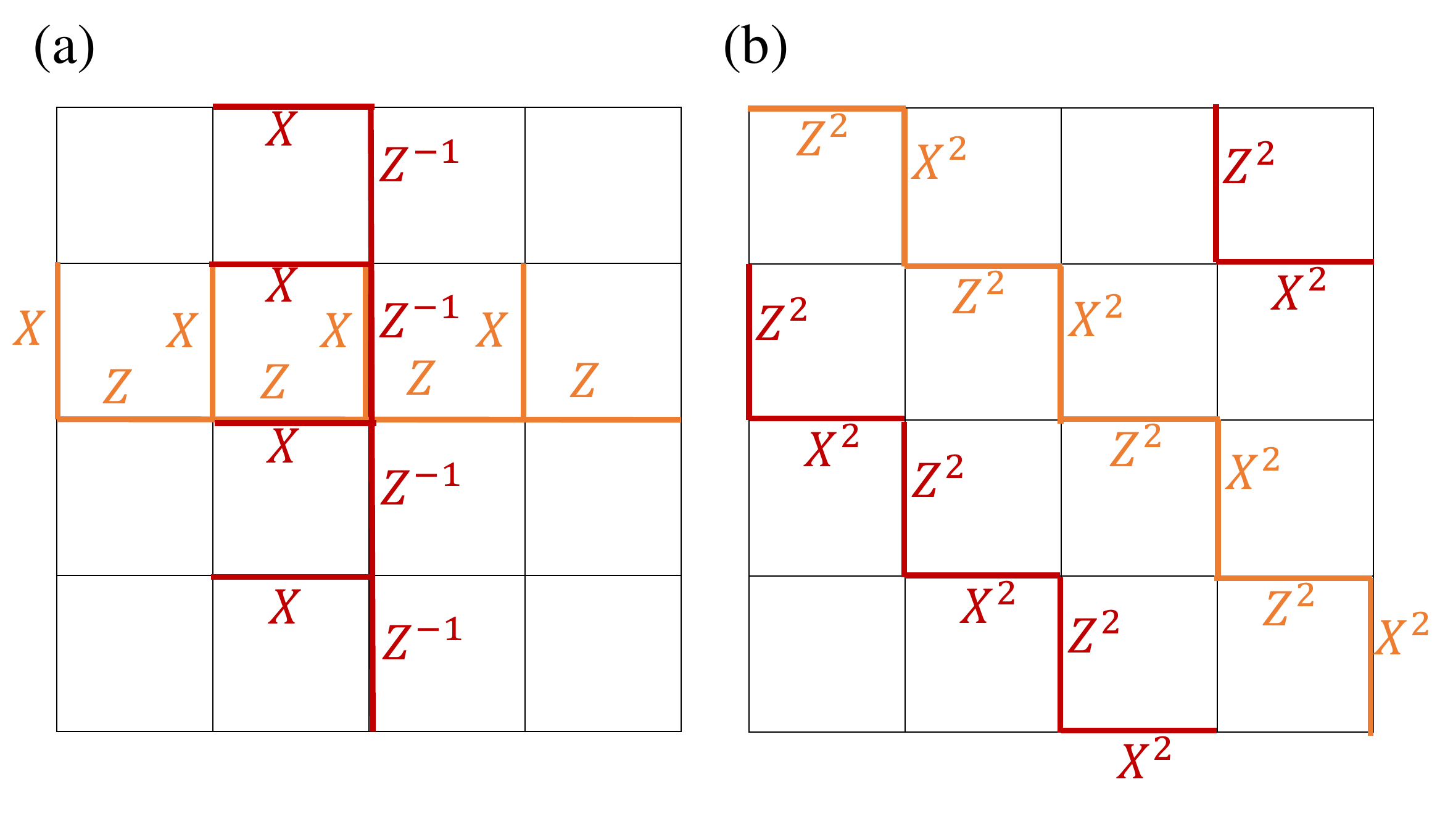}

\caption{(a). Two logical operators $W^{e^3m}_{\tilde{\gamma_x}}$ (in orange) and $W^{e^3m}_{\tilde{\gamma_y}}$ (in red). (b). Two non-local stabilizers. In both (a). and (b). periodic boundary conditions are imposed.}

\label{fig:nonlocal}
\end{figure}

\begin{figure}[htb]
 \centering
\includegraphics[width=1\linewidth]{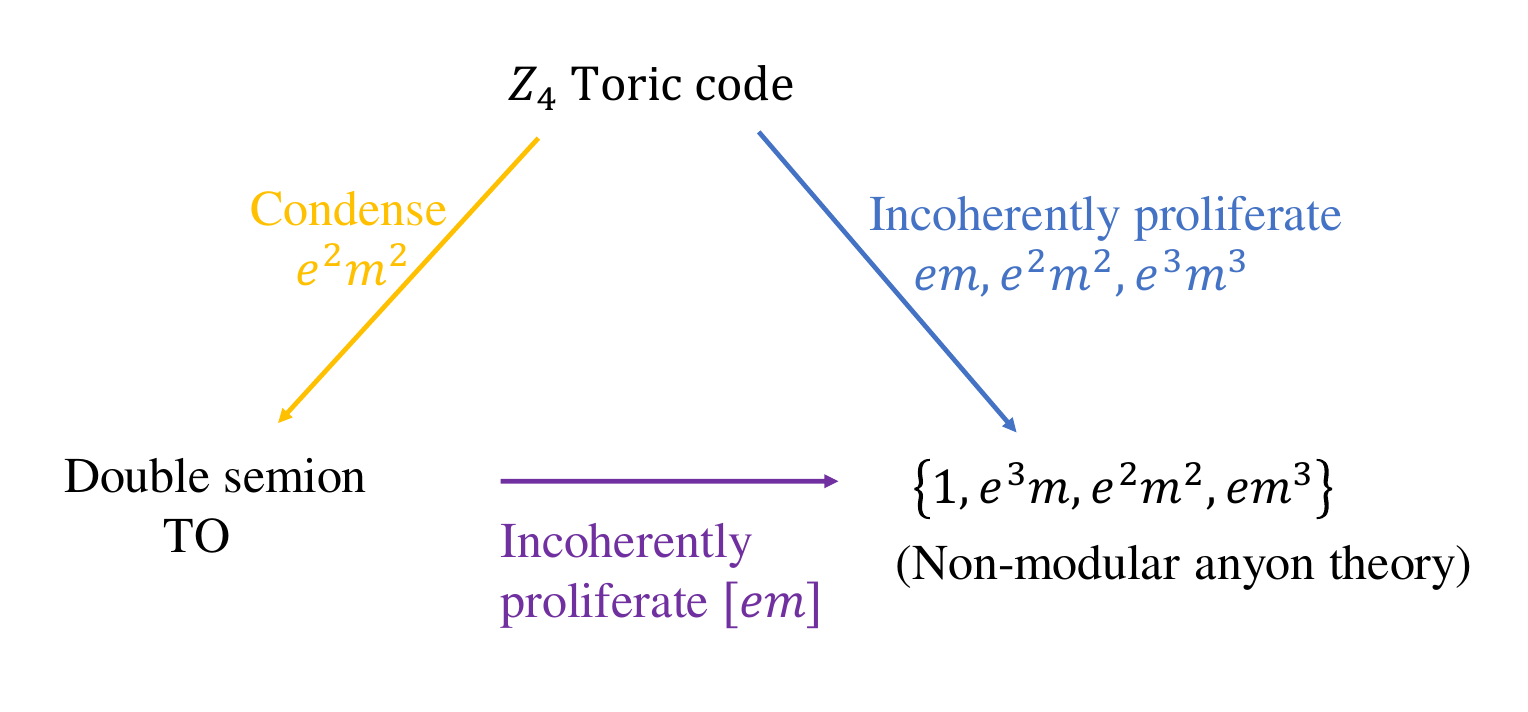}

\caption{Relations between $\Z_4$ toric code, double semion, and the non-modular anyon theory supported by intrinsic mixed-state TO.}

\label{fig:descendants}
\end{figure}

First, we note that the $1$-form symmetry generated by $e^3m$ is indeed preserved because $[W^{e^3m}_\gamma,K_i]=0$, for any closed $e^3m$ strings, which implies that $e^3m$ is a deconfined anyon. Examples of a shortest closed $e^3m$ string as well as an open $e^3m$ string is given in Fig.~\ref{fig:DS}(c). On the other hand, the $C_i$ terms in $H_{\text{DS}}$ do not commute with the Kraus operators. As a result, for any open $e^2m^2$ string $W^{e^2m^2}_C=\prod_{i\in C}C_i$, $\tr(\rho W_C^{e^2m^2})$ becomes $0$ for $\rho=\mathcal{N}^{[em]}[\rho_0]$. This means that the $e^2m^2$ anyons are revived from the Bose-Einstein condensate and become a detectable anyon. Notably, the $e^2m^2$ still proliferate classically, which is very different from Bose-Einstein condensation as we noted previously. Moreover, due to the strong $1$-form symmetry generated by $e^3m$ and the fusion rule $e^3m\times e^3m=e^2m^2$, $e^2m^2$ must become a deconfined anyon. Therefore, the remaining strongly deconfined anyons in $\rho$ form a $\Z_4$ group $\{1,e^3m,e^2m^2,em^3\}$.  We note that $e^3m$ and $em^3$ are both anti-semions and $e^2m^2$ is a transparent boson, meaning that 
its presence cannot be remotely detected via an Aharonov-Bohm measurement, i.e., a full braid of any strongly deconfined anyon around it only results in a unity phase factor. Anyon theories with transparent bosons/fermions are known as non-modular anyon theories \cite{kitaev2006anyons,Ellison2023subsystem}. It is widely believed that non-modular anyon theories cannot be realized by local gapped Hamiltonians in 2D bosonic systems \cite{levin2013protected,kong2014braided}. This implies the lack of a pure-state counterpart of the mixed-state TO; thus it is indeed intrinsically mixed. Notably, the intrinsic mixed-state TO constructed in Section \ref{sec:decoheredTC} is also non-modular, with strongly deconfined anyons $\{1,f\}$.  One crucial difference is that here the intrinsic mixed-state TO does have a quantum memory. The undamaged part of the stored information is manipulated by the logical operators shown in Fig.~\ref{fig:nonlocal}.
\begin{equation}
W^{e^3m}_{\tilde{\gamma_x}}=\prod_{i\in\tilde{\gamma_x}}X_iZ_{i+\bm{\delta}}, W^{e^3m}_{\tilde{\gamma_y}}=\prod_{i\in\tilde{\gamma_y}}X_iZ^{-1}_{i+\bm{\delta}}
\end{equation}
$W^{e^3m}_{\tilde{\gamma_x}}W^{e^3m}_{\tilde{\gamma_y}}=-W^{e^3m}_{\tilde{\gamma_y}}W^{e^3m}_{\tilde{\gamma_x}}$ as a consequence of the nontrivial self-braiding statistics of $e^3m$. Besides, both $(W^{e^3m}_{\tilde{\gamma_x}})^2,(W^{e^3m}_{\tilde{\gamma_x}})^2$ are elements of the stabilizer group defined by $H_{\text{DS}}$, thus acting trivially in the code space. Therefore, the intrinsic mixed-state TO supports quantum memory with one and only one logical qubit. 
Remarkably, the anyon theory here is identical to the one obtained by incoherently proliferating the $em,e^2m^2,e^3m^3$ in the $\Z_4$ toric code via the same channel $\mathcal{N}^{[em]}$ \cite{sohal2024noisy,cheng2024towards}. The relation between $\Z_4$ toric code, double semion, and the non-modular anyon theory is summarized in Fig.~\ref{fig:descendants}. However, it does not imply that the final mixed states in the two models are identical.
Actually, the decohered $\Z_4$ toric code model (under $\mathcal{N}^{[em]}$) is fully characterized by the stabilizer group $G_{e^3m}=\langle \{W^{e^3m}_p\}\rangle$, which defines the code space $\mathcal{H}_{C}$:
\begin{equation}
\mathcal{H}_C=\{|\psi\rangle,g|\psi\rangle=|\psi\rangle,\forall g\in G_{e^3m}\}
\label{eq:DS_codespace}
\end{equation}
In the maximally decohered case, the final state is the maximally mixed state in $\mathcal{H}_C$. For the decohered double semion model, however, there are additional non-local stabilizers formed by products of $C_i$. Two such non-local stabilizers are depicted in Fig.~\ref{fig:nonlocal}, and other non-local stabilizers can be obtained from these two via translation along the horizontal direction. 

We summarize several surprising features of intrinsic mixed-state TO revealed by this example. First, novel non-modular TO beyond the usual unitary modular tensor category description of 2+1D TO can be easily realized by Pauli stabilizer models under decoherence \cite{sohal2024noisy,cheng2024towards}. Secondly, decoherence can sometimes give rise to new types of deconfined anyons that is absent in the anyon theory supported by the ground-state TO. It further implies that some features of the mixed-state TO can go beyond the prediction based on the IR theory/anyon data of the original topological order (i.e., before decoherence), including the field-theoretic description \cite{lee2023quantum,bao2023mixed}. Indeed, if we start from other lattice realizations of double semion topological order, e.g., the $\Z_4$ Pauli stabilizer code introduced in \cite{Ellison2023subsystem} (which is a variant of \eqref{eq:H_DS}) or the twisted $\Z_2$ lattice gauge theory in \cite{Dijkgraaf1990,levin2012braiding,hu2013twisted,Fuente2021nonpaulitopological,song2024fracton}, and incoherently proliferate semions via quantum channels, we would get a chiral anti-semion theory $\{1,\bar s\}$ instead. The corresponding channels can be constructed by just taking the Kraus operators to be the semion string operators. For the explicit expression of the shortest string operators of twisted $\Z_2$ gauge theory, we refer the readers to the supplemental material of \cite{song2024fracton}.


\section{ Discussion\label{sec:discussion}}
Our work introduces a promising mechanism for creating novel topologically ordered phases in mixed states. We give two complementary perspectives to demonstrate such a possibility. The first perspective is to look at what anyons are proliferated. One of our key observations is that while the ways of anyon condensation are limited for pure states, anyon proliferation in mixed states can occur in more general ways, offering new possibilities for topological order. In the three models studied in this work, we propose new types of topological order arising from incoherent proliferation of fermionic or semionic anyons in ground-state topological order, which drives an unconventional phase transition that does not resemble any anyon condensation transition in pure states. The other perspective is to look at what remains. We find that the existence of anomalous strong
$1$-form symmetries can be viewed as a guiding principle to novel mixed-state TO. We give general proof that anomalous strong 1-form symmetries imply the LRE nature of the mixed states, which manifests as deconfined anyons with nontrivial statistics. By analyzing the fusion rules and statistics of deconfined anyons, we show the possibility of realizing non-modular TO in mixed states.

As is clear from the three examples, intrinsic mixed-state TO can have or not have quantum memory, depending on whether the remaining deconfined anyons have nontrivial braiding statistics or only fermionic statistics \cite{bao2023mixed}.  

It is worth noting that the construction of such exotic mixed states is experimentally feasible in current NISQ devices \cite{satzinger2021realizing}. For example, one can realize the decohered $\Z_2$ toric code model in Section \ref{sec:decoheredTC} by implementing incomplete error correction, where only the error syndrome with $A_vB_{p=v+\bm{\delta}}=-1$ is corrected after the syndrome measurement using string operators $W^e$ or $W^m$. This partial error correction would lead to a mixed state similar to $\rho_f$.

We end with some open directions. Firstly, as already mentioned in Section \ref{sec:generalities}, it remains unclear whether $1$-form symmetries generated by fermions also guarantee bipartite long-range entanglement, or whether there exist counterexamples that are bipartite separable but have multipartite long-range entanglement \cite{lessa2024mixed}. Secondly, a systematic classification of the intrinsic mixed-state TO is still lacking, with two main difficulties. The first is how to treat the weakly deconfined anyons with the peculiar one-way braiding statistics. The other is how to generalize the discussion of anyon statistics to mixed states away from the fixed-point models, like the one discussed in \ref{sec:robustness}. 
Finally, in our construction, we start from a topologically ordered state and get its descendants via noisy channels. It is also tempting to find systematic ways to prepare mixed-state topological order from short-range entangled mixed states, for example, by measuring mixed-state symmetry protected topological order \cite{ma2023average,lee2022symmetry,zhang2022strange,ma2023topological,lu2023mixed}.

\begin{acknowledgements}
{\it Acknowledgments}.---We are especially grateful to Ruihua Fan for many pieces of valuable advice and feedback on the manuscript. We also thank Tsung-Cheng Lu for pointing out to us the dependence of TEN on the choice of entanglement cut. We thank Yingfei Gu, Meng Cheng, Yu-An Chen, Zhen Bi, Jing-Yuan Chen, and He-Ran Wang for helpful discussions. This work is supported by the NSFC under Grant No. 12125405,  National Key R\&D Program of China (No. 2023YFA1406702), and the Innovation Program for Quantum Science and Technology (No. 2021ZD0302502). Z Wu acknowledges the support in part from Shuimu fellowships at Tsinghua University and the Engineering and Physical Sciences Research Council (EPSRC) fellowships.
\end{acknowledgements}

\appendix
\begin{widetext}
\section{Details about the decohered toric code \label{app:decoheredTC}}

\subsection{Calculation of the coherent information and mapping to the RBIM}
\label{cohe_inf}
In this section, we present detailed calculations of the coherent information $I_c=S(\rho_f)-S(\rho_{Rf})$, and the derivation of the mapping to the RBIM along the Nishimori line (\eqref{eq:Ic_mapping}) \cite{Nishimori1981internal}. In the next two subsections, we calculate the two von Neumann entropy $S(\rho_{f})$ and $S(\rho_{Rf})$ respectively, using the replica trick: $S=-\Tr(\rho\log{\rho})=-\lim_{n\rightarrow1}\frac{\partial }{\partial n}\Tr(\rho^n)$.



\subsubsection{von Neumann entropy $S(\rho_{Rf})$ }
We begin with the calculation of $S(\rho_{Rf})$. We introduce $2$ reference qubits, denoted by $\sigma_{1,2}$, and maximally entangle them with the two logical qubits in the ground state subspace of the system: 
\begin{equation}
 \ket{\Psi}= \frac{1}{2}\sum_{a,b=\pm 1}\ket{a,b}_S\otimes\ket{\sigma^z_1=a,\sigma^z_2=b}_R,
 \end{equation}
 where $a,b$ label the eigenvalues of the two non-contractible Wilson loops $W_{\gamma_x}^z,W_{\gamma_y}^z$ respectively, with the $U(1)$ phase ambiguity fixed by
\begin{equation}
    \ket{-1,1}_S=W_{\tilde{\gamma}_y}^x\ket{1,1}_S, \quad  \ket{1,-1}_S=W_{\tilde{\gamma}_x}^x\ket{1,1}_S, \quad \ket{-1,-1}_S=W_{\tilde{\gamma}_y}^xW_{\tilde{\gamma}_x}^x\ket{1,1}_S,
\end{equation}
where $W^{x}_{\tilde{\gamma}_{x,y}}$ are non-contractible $X$ loops on the dual lattice. It is straightforward to check that $|\Psi\rangle$ is a purification of $\rho_0$, the maximally mixed state in the ground-state subspace, $\rho_0=\text{tr}_R(|\Psi\rangle\langle\Psi|)$. One can alternatively view the reference qubits as an input (via the Choi map), which encodes information into the code space \cite{fan2023diagnostics}.

To facilitate the calculation of the coherent information, we write the decohered $\rho_{Rf}$ in the error chain representation: 
\begin{equation}
\rho_{Rf}=\sum_{C}P(C)W^f_C|\Psi\rangle\langle\Psi|
W^f_C
\label{eq:errorstring}
\end{equation}
where $C$ stands for error chain configurations (the set of links where error occurs) with total length $|C|$.  $P(C)=p^{|C|}(1-p)^{N-|C|}$ is  the occurrence probability of the error chain $C$ \footnote{For convenience, we drop the subscript of $p_f$ in Appendix \ref{cohe_inf} to \ref{weak_sym}}. $W^f_C$ is the (product of) open string operators which  create $f$ anyons at the ends of the $C$. 

Now the trace $\Tr(\rho_{Rf}^n)$ is,
\begin{equation}
\begin{aligned} 
\Tr(\rho_{Rf}^n)&= \sum_{\left\{C^{(s)}\right\}} \prod_{s=1}^{n} P\left(C^{(s)}\right) \operatorname{tr}\left[\prod_{s=1}^{n} \left(W^f_{C^{(s)}} |\Psi\rangle\langle\Psi|W^f_{C^{(s)}}\right) \right],\\
&=\sum_{\left\{C^{(s)}\right\}}\prod_{s=1}^{n} P\left(C^{(s)}\right)\left\langle\Psi\left|W^f_{C^{(s)}} W^f_{C^{(s+1)}}\right| \Psi\right\rangle,
\end{aligned}
\end{equation}
where $W^f_{C^{(n+1)}}\equiv W^f_{C^{(1)}}$ and the loops $C^{(s)}$ satisfy
\begin{equation}
C^{(s+1)}=C^{(1)}+\partial v^{(s)}, \quad s=1,2, \ldots, n-1
\end{equation}
to give nonzero contribution. $\partial v^{(s)}$ are boundaries of a set of plaquettes $v^{(s)}$, so they are homologically trivial loops. Then $\Tr(\rho_{Rf}^n)$ can be further simplified as
\begin{equation}
\Tr(\rho_{Rf}^n)=\frac{1}{2^{n-1}}\sum_{C^{(1)}} P\left(C^{(1)}\right) \sum_{\left\{v^{(s)}\right\}} \prod_{s=1}^{n-1} P\left(C^{(1)}+\partial v^{(s)}\right).
\end{equation}
The prefactor $\frac{1}{2^{n-1}}$ is due to the fact that for each replica $s=1,2,\cdots n-1$, there are two plaquette sets $v^{(s)}$ giving the same boundary $\partial v^{(s)}$. $\Tr(\rho_{Rf}^n)$ can be mapped to the partition function of a classical Ising model with $n-1$ flavors of Ising spin and a defect line at $C^1$. Concretely, we introduce $Z_2$ variables $n_{v^{(s)}}(l)=1,0$ to denote whether link $l$ is occupied in $\partial v^{(s)}$ or not. Then we can express the probability $P\left(C^{(1)}+\partial v^{(s)}\right)$  by the $Z_2$ variables $n_{v^{(s)}}(l)$. For example, if a link $l\in C^{(1)}$ and $n_{v^{(s)}}(l)=1$, then link $l$ does not occur in the error chain $C^{(1)}+\partial v^{(s)}$ and contributes a factor $(1-p)^{n_{v^{(s)}}(l)}p^{1-n_{v^{(s)}}(l)}$ in $P\left(C^{(1)}+\partial v^{(s)}\right)$.  As a result, the probability $P\left(C^{(1)}+\partial v^{(s)}\right)$  can be written as
\begin{equation}
\begin{aligned}
    &P\left(C^{(1)}+\partial v^{(s)}\right)=\left[\Pi_{l\in C^{(1)}}\left( (1-p)^{n_{v^{(s)}}(l)} p^{1-n_{v^{(s)}}(l)}\right)\right]\left[\Pi_{l\notin C^{(1)}}\left( p^{n_{v^{(s)}}(l)} (1-p)^{1-n_{v^{(s)}}(l)}\right)\right].
\end{aligned}
\end{equation}
The first part with those links belonging to the error chain $C^{(1)}$ can be made symmetric as
\begin{equation}
\begin{aligned}
    \Pi_{l\in C^{(1)}}\left( (1-p)^{n_{v^{(s)}}(l)} p^{1-n_{v^{(s)}}(l)}\right)&=\Pi_{l\in C^{(1)}}\left(\sqrt{p(1-p)}(\frac{1-p}{p})^{n_{v^{(s)}}(l)-\frac{1}{2}}\right)\\
    &=\sqrt{p(1-p)}^{|C^{(1)}|}\Pi_{l\in C^{(1)}}(\frac{1-p}{p})^{n_{v^{(s)}}(l)-\frac{1}{2}}
    \end{aligned}.
\end{equation}
Similarly, we also make the second part symmetric as
\begin{equation}
\begin{aligned}
    \Pi_{l\notin C^{(1)}}\left( p^{n_{v^{(s)}}(l)} (1-p)^{1-n_{v^{(s)}}(l)}\right)&=\Pi_{l\notin C^{(1)}}\left(\sqrt{(1-p)p}(\frac{p}{1-p})^{n_{v^{(s)}}(l)-\frac{1}{2}}\right)\\
    &=\sqrt{p(1-p)}^{(N-|C^{(1)}|)}\Pi_{l\notin C^{(1)}}(\frac{p}{1-p})^{n_{v^{(s)}}(l)-\frac{1}{2}}.
\end{aligned}
\end{equation}
Then we can express the link probability part $(\frac{p}{1-p})^{n_{v^{(s)}}(l)-\frac{1}{2}}$ or $(\frac{1-p}{p})^{n_{v^{(s)}}(l)-\frac{1}{2}}$ as an Ising coupling between two nearest-neighbour plaquettes which share the link $l$. Concretely, we introduce $n-1$ flavours of Ising spins $\tau^{(s)}=\pm1,s=1,2,...n-1$ on each plaquette, and introduce the Ising coupling constant $J$ as: $e^{-2 J}=p /(1-p)$. Then the link probability part $(\frac{p}{1-p})^{n_{v^{(s)}}(l)-\frac{1}{2}}$ or $(\frac{1-p}{p})^{n_{v^{(s)}}(l)-\frac{1}{2}}$ can be written as $\exp[J\eta_{ij}\tau_i^{(s)}\tau_j^{(s)}]$, where $i,j$ are the dual lattice site coordinates of the two plaquettes, and $\eta_{ij}=-1\ (1)$ for $l$ belonging to (not belonging to) the error chain $C^{(1)}$. Then $p$ is the probability of antiferromagnetic coupling  for each bond.

As a result, $\Tr(\rho_{Rf}^n)$ can be expressed as the partition function of a random bond Ising model (RBIM) with $n-1$ flavours of Ising spins and periodic boundary condition (PBC).:
\begin{equation}
\begin{aligned}
    \Tr(\rho_{Rf}^n)&=\frac{1}{2^{n-1}}\sum_{C^{(1)}} P\left(C^{(1)}\right) \sum_{\left\{v^{(s)}\right\}} \prod_{s=1}^{n-1} P\left(C^{(1)}+\partial v^{(s)}\right)\\
    &=\frac{1}{2^{n-1}}\left(\sqrt{(1-p)p}\right)^{(n-1)N}\sum_{C^{(1)}} P\left(\{\eta\}\right) \sum_{\left\{\tau^{(s)}\right\}} \prod_{s=1}^{n-1} \exp[J\eta_{ij}\tau_i^{(s)}\tau_j^{(s)}]\\
    &=\frac{1}{2^{n-1}}\left(\sqrt{(1-p)p}\right)^{(n-1)N}\sum_{C^{(1)}} P\left(\{\eta\}\right)\prod_{s=1}^{n-1}\sum_{\left\{\tau^{s}\right\}}\exp[J\eta_{ij}\tau_i^{(s)}\tau_j^{(s)}]\\
    &=\frac{1}{2^{n-1}}\left(\sqrt{(1-p)p}\right)^{(n-1)N}\sum_{C^{(1)}} P\left(\{\eta\}\right)\left(Z[J,\{\eta\}]\right)^{n-1}.
\end{aligned}
\end{equation}
Finally we take the replica limit $n\rightarrow 1$ to derive the von Neumann entropy $S(\rho_{Rf})$:
\begin{equation}
\begin{aligned}
   S(\rho_{Rf})&=- \lim_{n\rightarrow1}\frac{\partial }{\partial n}\Tr(\rho_{Rf}^n)\\
   &=-\frac{N}{2}\log [p(1-p)]+\log 2-\sum_{\{\eta_l\}}P(\{\eta\})\log Z[J,\{\eta\}]\\
   &\equiv -\overline{\log Z^{\text{RBIM}}_{\text{PBC}}}+\log2-\frac{N}{2}\log [p(1-p)] ,
\end{aligned}
\end{equation}
where the first term is the average free energy of the RBIM along the Nishimori line: $e^{-2J}=\frac{p}{1-p}$.
\subsubsection{von Neumann entropy $S(\rho_{f})$ }
The von Neumann entropy $S(\rho_{f})$ can be derived similar to $S(\rho_{Rf})$. The initial density matrix is 
\begin{equation}
    \rho_{0}=\frac{1}{4}\sum_{a,b=\pm 1}\ket{a,b}\bra{a,b}.
\end{equation}
Then $ \Tr(\rho_{f}^n)$ is
\begin{equation}
\begin{aligned} 
\Tr(\rho_{f}^n)&=\sum_{\left\{C^{(s)}\right\}}\sum_{a^{(s)},b^{(s)}}\prod_{s=1}^{n} P\left(C^{(s)}\right)\left(\frac{1}{4}\bra{a^{(s)},b^{(s)}}W^f_{C^{(s)}} W^f_{C^{(s+1)}} \ket{a^{(s+1)},b^{(s+1)}}\right)\\
&=\frac{1}{2^{n-1}}\cdot\frac{1}{4^{n-1}}\sum_{C^{(1)}} P\left(C^{(1)}\right) \prod_{s=1}^{n-1} \sum_{\left\{v^{(s)}\right\}}\sum_{d_x^{(s)},d_y^{(s)}=0,1} P\left(C^{(1)}+\partial v^{(s)}+d_x^{(s)}\gamma_x+d_y^{(s)}\gamma_y\right),
\end{aligned}
\label{zhenji2}
\end{equation}
where $n+1\equiv 1$, and $d^{(s)}_{x,y}=0,1$ denotes whether or not $C^{(s)}$ lies in the same homological class as $C^{(1)}$. 
Similar to the mapping of $\Tr(\rho_{Rf}^n)$ to the partition function of RBIM, we can also map $\Tr(\rho_{f}^n)$ to the  partition function of RBIM, except that here we must sum over the four contributions of inserting or not the two non-contractible defect lines on the torus:
\begin{equation}
\begin{aligned}
\Tr(\rho_{f}^n)&=\frac{1}{2^{n-1}}\cdot\frac{1}{4^{n-1}}\left(\sqrt{(1-p)p}\right)^{(n-1)N}\sum_{C^{(1)}} P\left(\{\eta\}\right)\left(\sum_{d_{x},d_{y}=0,1} Z_{d_x,d_y}[J,\{\eta\}]\right)^{n-1}\\
&\equiv \frac{1}{2^{n-1}}\cdot\frac{1}{4^{n-1}}\left(\sqrt{(1-p)p}\right)^{(n-1)N}\overline{\left(\sum_{d_x,d_y=0,1}Z^{\text{RBIM}}_{d_x,d_y}\right)^{n-1}}
\end{aligned}
\label{eq:RBIM2}
\end{equation}
where $Z^{\text{RBIM}}_{d_x,d_y}$ is the partition function with $d_a$ non-contractible defect lines inserted along the cycle $\gamma_a$. Along the defect line the coupling changes from $\eta J$ to $-\eta J$. This is equivalent to taking the anti-periodic boundary condition (APBC). 

$S_{\rho_f}$ can in turn be obtained by taking the replica limit:
\begin{equation}
S(\rho_f)=- \lim_{n\rightarrow1}\frac{\partial }{\partial n}\Tr(\rho_{f}^n)=3\log 2-\overline{ \log \left[\sum_{d_x,d_y=0,1 }Z^{\text{RBIM}}_{d_x,d_y}\right]     }-\frac{N}{2}\log[p(1-p)].
\end{equation}
We note that the second term can also be understood as the free energy of RBIM with all four types of boundary condition (PBC/APBC along $x,y$ direction) into account. 

\subsubsection{Critical error rate and classical memory from coherent information }
As we have demonstrated in the previous subsections, $S_{\rho_{f}}$ and $S_{\rho_{Rf}}$ can be mapped to the free energy of the RBIM with or without the insertion of non-contractible defect lines (plus some constants),  so the coherent information $I_c$ is related the excess free energy of the defect line,

\begin{equation}
I_c=2\log 2-\overline{\log \frac{\sum_{d_x,d_y=0,1 }Z^{\text{RBIM}}_{d_x,d_y}}{Z^{\text{RBIM}}_{00}}}=2\log 2-\overline{\log\left[\sum_{d_x,d_y=0,1 }e^{-\Delta F_{d_x,d_y}}\right]},
\end{equation}
where $\Delta F_{d_x,d_y}$ is the excess free energy with the insertion of a non-contractible defect line. For small $p$, the RBIM is in the ferromagnetic (FM) phase, and the excess free energy of a defect line is extensive, $\Delta F_{\{d_x,d_y\}\neq\{0,0\}}\sim O(L)$, which leads to $I_c=2\log 2$. On the other hand, when $p$ is above the error threshold $p_c\approx 0.109$, the RBIM undergoes a phase transition to a paramagnetic (PM) phase, and $I_c $ drops to $0$ in the thermodynamic limit, which indicates that $\rho_f$ only retains a classical memory. This is exactly the same as the situation with single-qubit errors. 

\subsection{Relative entropy}
As mentioned in the main text, the phase transition at $p_f=p_c$ is driven by the proliferation of $f$ anyons. In the double space, this corresponds to the condensation of $f_+f_-$. In this section we provide a quantitative diagnosis of the $f$ anyon proliferation in the original Hilbert space. We denote the string operators creating $\alpha$ anyons at the ends of the string as $w^\alpha$, and investigate whether $\rho^\alpha_{f}\equiv\mathcal{N}^f[w^\alpha\rho_0 w^\alpha]$ is really a distinct state from $\rho_f$. Quantitatively, we calculate the relative entropy:
\begin{equation}
D(\rho_f||\rho_f^\alpha)\equiv\Tr(\rho_f\log\rho_f)-\Tr(\rho_f\log\rho_f^\alpha),
\end{equation}
and examine whether it diverges as the length of $w^\alpha$ approaches infinity, which is proposed as a generalization of Fredenhagen-Marcu order parameter for ground states \cite{fan2023diagnostics,fredenhagen1983charged,gregor2011diagnosing}.

It turns out that, for $p_f<p_c$, $D(\rho_f||\rho_f^\alpha)$ diverges for all three types of anyons, while for $p_f>p_c$, $D(\rho_f||\rho_f^f)$ becomes finite, in agreement with our expectation. Additionally, although $D(\rho_f||\rho_f^{e(m)})$ is divergent, $e,m$ cease to be distinct (weakly) deconfined excitations since $e\times f=m$.

To obtain $D(\rho_f||\rho_f^\alpha)$, we can still use the replica trick:
\begin{equation}
D^{(n)}(\rho_f||\rho^\alpha_f)\equiv\frac{1}{1-n}\log\frac{\Tr\rho_f(\rho_f^\alpha)^{n-1}}{\Tr\rho_f^n},
\end{equation}
and revover $D(\rho_f||\rho^\alpha_f)$ by taking the limit $n\rightarrow 1$. 

Using the error chain expansion, we get
\begin{equation}
\begin{aligned} 
\Tr\rho_{f}(\rho_f^\alpha)^{n-1}&=\sum_{\{C^{(s)}\}}\prod_{s=1}^{n}P(C^{(s)})\Tr\left(W^f_{C^{(1)}}\rho_0 W^f_{C^{(1)}}\prod_{s=2}^n W^f_{C^{(s)}}w^\alpha\rho_0 w^\alpha W^f_{C^{(s)}}\right)\\
&=\sum_{\left\{C^{(s)}\right\}}\sum_{a^{(s)},b^{(s)}}\left[\prod_{s=1}^{n} \frac{1}{4}P\left(C^{(s)}\right)\right]\bra{a^{(1)},b^{(1)}}W^f_{C^{(1)}}W^f_{C^{(2)}}w^\alpha\ket{a^{(2)},b^{(2)}}\bra{a^{(n)},b^{(n)}}w^\alpha W^f_{C^{(n)}} W^f_{C^{(1)}} \ket{a^{(1)},b^{(1)}}\\
&\qquad\qquad\qquad\ \prod_{s=2}^{n-1}\bra{a^{(s)},b^{(s)}}w^\alpha W^f_{C^{(s)}}  W^f_{C^{(s+1)}}w^\alpha \ket{a^{(s+1)},b^{(s+1)}}.
\end{aligned}
\end{equation}
Clearly, for $\alpha=e,m$ , $\Tr\rho_{f}(\rho_f^\alpha)^{n-1}=0$ , so the relative entropy diveges. Thus we only focus on $\alpha=f$ next. Terms in the summation is nonvanishing only if the error chain configurations satisfy the following condition:
\begin{equation}
C^{(s)}=C^{(1)}+\partial v^{(s)}+d^{(s)}_x\gamma_x+d^{(s)}_y\gamma_y+A,\ s=1,2,\cdots ,n-1,
\end{equation}
where $d_{x/y}^{(s)}=0,1$ and $A$ denotes the string where $w^{\alpha=f}$ acts nontrivially. Compared to \eqref{zhenji2}, \eqref{eq:RBIM2}, we can see that the insertion of $w^f_A$ corresponds to inserting an additional defect line along $A$ in the RBIM, which means the Ising coupling flips sign along $A$. Denote the partition function of RBIM with defect line along $A$ as $Z^{\text{RBIM}}[A]$, where we implicitly sum over the four types of boundary conditions $\{d_x,d_y\}$. Then
\begin{equation}
D^{(n)}(\rho_f||\rho^f_f)=\frac{1}{1-n}\log\frac{\overline{(Z^{\text{RBIM}}[A])^{n-1}}}{\overline{(Z^{\text{RBIM}})^{n-1}}},
\end{equation}
Taking the replica limit $n\rightarrow 1$, we obtain the relative entropy:
\begin{equation}
D(\rho_f||\rho_f^f)=\overline{\log Z^{\text{RBIM}}}-\overline{\log Z^{\text{RBIM}}[A]},
\end{equation}
which is mapped to the excess free energy of defect line $A$. In the FM phase $(p<p_c)$, it diverges as the distance between the two ends of $A$ goes to infinity. However, in the PM phase $(p>p_c)$, it is finite, which indicates the incoherent proliferation of $f$.

\subsection{Calculation of the entanglement negativity}

In this section, we derive the entanglement negativity $\varepsilon_A(\rho_f)\equiv\log ||\rho^{T_A}_f||_1$. It turns out that to calculate $\varepsilon_A(\rho_f)$, it is more convenient to use the loop expansion (\eqref{loop} in the main text) instead of the error chain expansion used in the last two sections. 
We start from \eqref{tension} in the main text:

\begin{equation}
\rho_{f}^{T_A}=\frac{1}{2^N}\sum_{g\in G}(1-2p)^{l_g}y_A(g)g,
\end{equation}
where $l_g$ is the length of the segment where $g_x$ and $g_z$ does not coincidence, and
$$
y_A(g)\equiv \sign_A(g_x,g_z)\equiv
\left\{
\begin{array}{ll}
1,\quad &\text{if } g_{xA},g_{zA} \text{ commute.} \\
-1,\quad &\text{if } g_{xA}, g_{zA} \text{ anticommute.}
\end{array}
\right.
$$  
 To calculate $\varepsilon_A(\rho_f)$, we utilize the replica trick, that is, we first calculate the $2n^{th}$ Renyi negativity $\varepsilon^{(2n)}_A(\rho_f):=\frac{1}{2-2n}\log \frac{\Tr(\rho^{T_A}_f)^{2n}}{\Tr\rho_f^{2n}}$ and finally take the replica limit $2n\rightarrow 1$.

First, we consider the bipartition of $A\bigcup \bar A$ of a cylinder, as shown in Fig.~\ref{fig:fig1} (g)-(i) in the main text. We denote the bipartition in Fig.~\ref{fig:fig1}(h) as bipartition 1 and that in Fig. 1(i) as bipartition 2. We discuss these two types of bipartitions in detail below.

{\it Bipartition 1, generic $p$}. We deal with bipartition 1 first. In this case we are able to obtain an exact result of $\varepsilon_A^{(2n)}$ for any $p$. As we will show below, the result actually does not depend on $p$ at all.
To start with, we calculate $(\rho^{T_A}_f)^2$,
\begin{equation}
\begin{aligned}
(\rho^{T_A}_f)^2 &=\frac{1}{2^{2N}}\sum_{g,h\in G}(1-2p)^{l_g+l_h}y_A(g)y_A(h)gh\\
&=\frac{1}{2^{2N}}\sum_{g,h\in G}(1-2p)^{l_g+l_h}y_A(gh)\sign_A(g,h)gh\\
&=\frac{1}{2^{2N}}\sum_{g,\tilde{g}\in G}(1-2p)^{l_g+l_{g\tilde{g}}}y_A( \tilde{g})\sign_A(g,\tilde{g})\tilde{g}.
\end{aligned}
\label{eq:rhoT2}
\end{equation}
In the last step we use the substitution $h=g\gtd$. 

To simplify the expression, we deal with the summation over $g$ first. The crucial part in this expression is $\sign_A(g,\gtd)$ which leads to complete destructive inteference when $\gtd$ crosses the boundary between $A,\bar A$. To be more precise, we define the subgroup $H$ of $G$:
\begin{equation}
H\equiv\{g\in G|g_Ag^\prime_A=g^\prime_Ag_A,\forall g^\prime\in G\}.
\end{equation}
For bipartition 1, $H$ contains all loops that do not cross the boundary. 
Then we can simplify  \eqref{eq:rhoT2} :
\begin{equation}
(\rho^{T_A}_f)^2=\frac{1}{2^{2N}}\sum_{g\in G,\tilde{g}\in H}(1-2p)^{l_g+l_{g\tilde{g}}}y_A( \tilde{g})\tilde{g}.
\label{eq:en0}
\end{equation}
Thus

\begin{equation}
\begin{aligned}
\Tr(\rho^{T_A}_f)^{2n}&=2^{-2nN}\prod_{s=1}^{n}\sum_{\gtd^{(s)}\in H}\prod_{s=1}^{n}\sum_{g^{\textcolor{red}{(s)}}\in G}(1-2p)^{\sum_{s=1}^{n}l_{g^{(s)}}+l_{g^{(s)}\gtd^{(s)}} } y_A(\prod_{s=1}^n \gtd^{(s)})\Tr\left(\prod_{s=1}^{n}\gtd^{(s)}\right)\\
&=2^{(1-2n)N}\sum_{\gtd^{(1)},\cdots,\gtd^{(n-1)}\in H} \sum_{g^{(1)},\cdots,g^{(n)}\in G}(1-2p)^{ \sum_{s=1}^{n-1}(l_{g^{(s)}}+l_{g^{(s)}\gtd^{(s)}}) +l_{\prod_{s=1}^{n-1}g^{(s)} }+l_{\prod_{s=1}^{n-1}g^{(s)}\gtd^{(s)} } }.
\end{aligned}
\end{equation}

In a similar manner we can obtain $\Tr\rho_f^{2n}$, resulting in a similar expression with the summation over $H$ replaced by a summation over $G$:
\begin{equation}
 \Tr\rho_f^{2n}=2^{(1-2n)N}\sum_{\gtd^{(1)},\cdots,\gtd^{(n-1)}\in G} \sum_{g^{(1)},\cdots,g^{(n)}\in G}(1-2p)^{ \sum_{s=1}^{n-1}(l_{g^{(s)}}+l_{g^{(s)}\gtd^{(s)}}) +l_{\prod_{s=1}^{n-1}g^{(s)} }+l_{\prod_{s=1}^{n-1}g^{(s)}\gtd^{(s)} } }.   
\end{equation}
Since we are only concerned about their ratio, we can extract the common part in the expression and rename it as $O_{ \{\gtd\} }$.
\begin{equation}
O_{\{ \gtd \}  }=2^{(1-2n)N} \sum_{g^{(1)},\cdots,g^{(n)}\in G}(1-2p)^{ \sum_{s=1}^{n-1}(l_{g^{(s)}}+l_{g^{(s)}\gtd^{(s)}}) +l_{\prod_{s=1}^{n-1}g^{(s)} }+l_{\prod_{s=1}^{n-1}g^{(s)}\gtd^{(s)} } }.
\end{equation}
It is straightforward to show that $O_{\{ \gtd \}  }$ is actually only a function of $l_{\gtd^{(s)}}$. 
Based on this observation, we can divide the summation over $\{\gtd\}$ into different classes. First, we define the invariant subgroup $G_f$ of $G$, generated by $A_{v=p-\bm{\delta}}B_{p}$. In other words, elements in $G_f$ are tensionless loops with $l_g=0$. Similarly, we define the subgroup $H_f$ of $H$ to be $H_f\equiv\{g\in G_f|g_Ag^\prime_A=g^\prime_Ag_A,\forall g^\prime\in G_f\}$. Then
\begin{equation}
\begin{aligned}
\Tr(\rho_f^{T_A})^{2n}&=\sum_{\gtd^{(1)}_f,\cdots \gtd^{(n-1)}_f\in H_f}\sum_{\tilde{u}^{(1)},\cdots,\tilde{u}^{(n-1)}\in H/H_f } O_{\{\tilde{u}\}},\\
\Tr\rho_f^{2n}&=\sum_{\gtd^{(1)}_f,\cdots \gtd^{(n-1)}_f\in G_f}\sum_{\tilde{u}^{(1)},\cdots,\tilde{u}^{(n-1)}\in G/G_f } O_{\{\tilde{u}\}}
\end{aligned}
\label{eq:en1}
\end{equation}
Since $G/G_{f}=H/H_{f}$, we get 
\begin{equation}
\frac{  \Tr(\rho_f^{T_A})^{2n}\ } { \Tr\rho_f^{2n}  }=\left(\frac{|H_f| } { |G_f|   } \right)^{n-1}=2^{(2-2n)(L-1)},
\label{eq:en2}
\end{equation}
where $L$ is the length of the entanglement cut. Thus the Renyi negativity is:
\begin{equation}
\varepsilon^{(2n)}_A(\rho_f)=(L-1)\log 2,\forall n.
\end{equation}
In the replica limit $2n\rightarrow 1$, we get $\varepsilon_A(\rho_f)=(L-1)\log 2$. The subleading term, $\log 2$, is the topological entanglement negativity (TEN), which takes exactly the same value as that of the toric-code ground state.

{\it Bipartition 2, $p=\frac{1}{2}$}. For bipartition 2, the calculation for generic $p$ is much more challenging. This is mainly because $G/G_f\neq H/H_f$, which means \eqref{eq:en2} cannot be derived from \eqref{eq:en1} for generic $p$. As a result, the negativity does depend on $p$ for this bipartition. Here we are mainly interested in the phase with intrinsic mixed-state TO for $p>p_c$, so we take the maximally decohered limit, $p=\frac{1}{2}$, in which case the calculation can be greatly simplified. 

For $p=\frac{1}{2}$, 
\begin{equation}
\rho_f=\frac{1}{2^N}\sum_{g_f\in G_f}g_f=\frac{1}{2^{N/2+1}}\prod_p \frac{1+W_p}{2},\quad W_p:=A_{p-\bm{\delta}}B_{p}.
\end{equation}

In this case the negativity exhibits an unusual dependence on the parity of $L$, as shown in \eqref{eq:TEN_rough} in the main text. We will derive this result below. 

Analogous to \eqref{eq:rhoT2} and \eqref{eq:en0}, we have,
\begin{equation}
\begin{aligned}
(\rho^{T_A}_f)^2&=\frac{1}{2^{2N}}\sum_{g_f,\tilde{g}_f\in G_f}y_A( \tilde{g}_f)\sign_A(g_f,\tilde{g}_f)\tilde{g}_f\\
&=\frac{1}{2^{2N}}\sum_{g_f\in G_f,\tilde{g}_f\in H_f}y_A( \tilde{g}_f)\tilde{g}_f.
\end{aligned}
\label{eq:en3}
\end{equation}
Then it is straightforward to obtain the Renyi negativity:
\begin{equation}
\varepsilon^{(2n)}_A(\rho_f) =\frac{1}{2-2n}\log \left(\frac{|H_f| } { |G_f|   } \right)^{n-1}.
\end{equation}
For odd $L$, $\frac{|G_f|}{|H_f|}$ amounts to the number of elements in $G_f$ acting on the boundary: $\frac{|G_f|}{|H_f|}=2^{L-1}$. Here the ``$-1$'' is due to the fact that $h=\prod_{p_i \text{ on the boundary}}W_{p_i}$ is an element in $H_f$. Consequently, $\varepsilon^{(2n)}_A(\rho_f)=\frac{L-1}{2}\log2$. The negativity is half the value for bipartition 1, because there are only a half number of $W_{p}$ acting on the boundary. 

For even $L$, the calculation is subtler. Specifically, there exist special elements ${h_1}=\prod_{i=1,3,\cdots, L-1}W_{p_i}$, $h_2=\prod_{i=2,4,\cdots,L}W_{p_i}(=h\cdot h_1)$, that acts nontrivially on the boundary, but still belongs to $H_f$. See Fig.~\ref{fig:bipartitions}(a) for an illustration. Therefore, $\frac{|G_f|}{|H_f|}=2^{L-2},\varepsilon^{(2n)}_A(\rho_f)=\frac{L-2}{2}\log2$.

Taking the replica limit $2n\rightarrow 1$, we get the final result of logarithmic negativity (for bipartition 2):
\begin{equation}
\varepsilon_A(\rho_f)=
\left\{
\begin{aligned}
\frac{L}{2}\log 2-\log2,\quad  & \text{if } L \text{ is even,} \\ 
\frac{L}{2}\log 2-\frac{\log2}{2},\quad  & \text{if } L \text{ is odd.}
\end{aligned}
\right. 
\label{eq:bipartition2}
\end{equation}

In all the cases, we get a nonzero TEN, indicating nontrivial topological order. Here the value of TEN exhibits an unusual dependence on the parity of boundary size, and thus seems to be less universal than we would expect for topologically ordered phase. In the next section we will try to resolve this puzzle by relating entanglement properties of $\rho_f$ to those of more familiar ground-state topological order.

\begin{figure}[htb]
 \centering
\includegraphics[width=0.6\linewidth]{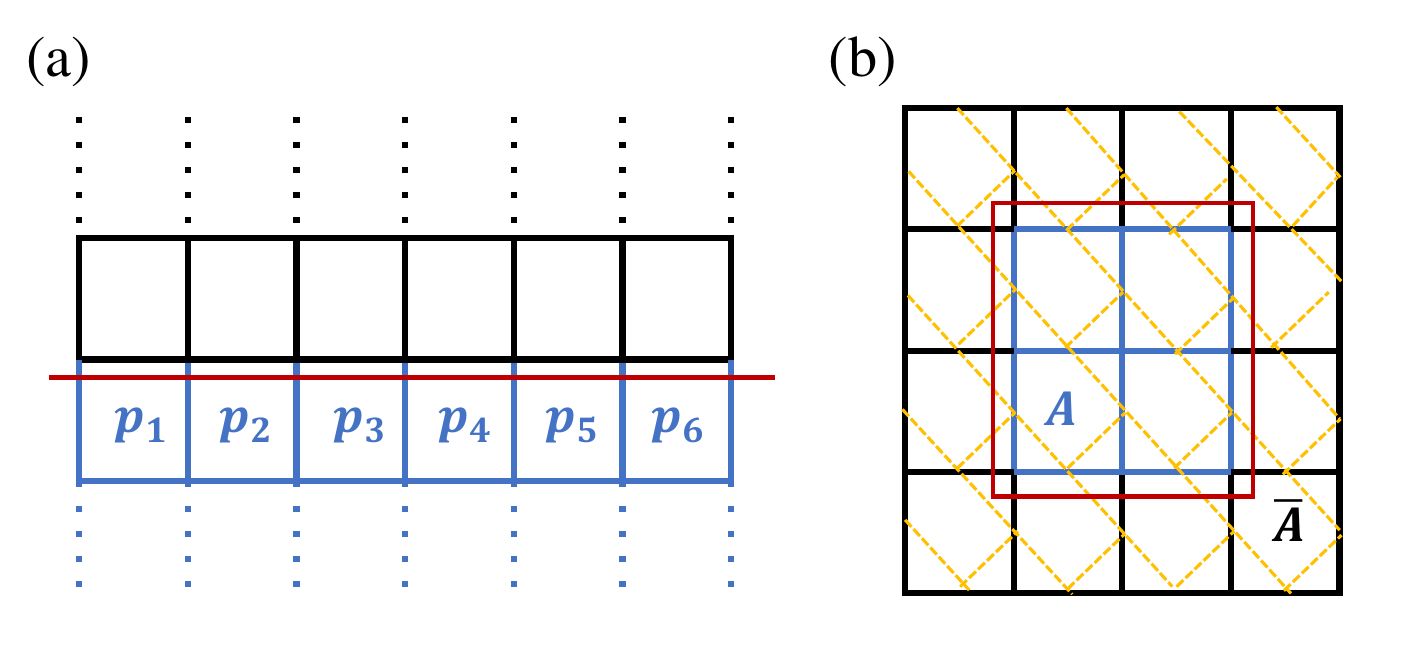}

\caption{(a). Bipartition 2 on a cylinder, $L=6$ in this case. (b). An example of bipartition with contractible subregion $A$. Orange dashed lines represent auxiliary links. $N_l=12$ in this case. }

\label{fig:bipartitions}
\end{figure}

Before that, we first make some quick comments on entanglement negativity for more generic bipartitions, including the cases with contractible subregion $A$. For $p=\frac{1}{2}$, we can perform similar calculations above and get exact results of negativity with general bipartitions. It is convenient to introduce some auxiliary links connecting the six qubits acted upon by $W_p$ for each $p$. We give one example in Fig.~\ref{fig:bipartitions}(b). With the help of these auxiliary links, the entanglement negativity for general bipartitions (with the only assumption that both $A$ and $\bar A$ are connected) yields,
\begin{equation}
\varepsilon_A(\rho_f)=
\left\{
\begin{aligned}
\frac{N_l}{2}\log 2-\log2,\quad  & \text{if } N_l \text{ is even,} \\ 
\frac{N_l}{2}\log 2-\frac{\log2}{2},\quad  & \text{if } N_l \text{ is odd.}
\end{aligned}
\right. 
\label{eq:bipartition2}
\end{equation}
$N_l$ counts the number of auxiliary links that are cut through by the entanglement cut. The even/odd dependence shows up again. The reason that bipartition 1 is free of this problem is that $N_l=2L$ is always even in that case. Finally, we note that due to the unusual dependence of TEN (defined as the value of the subleading term here) on the boundary size, one cannot extract TEN by calculating the tripartite mutual information as in the Kitaev-Preskill scheme, unless carefully choosing the multipartition such that $N_l$ have the same parity for all subregions. 
\subsection{Relation to translation-symmetry-enriched $\Z_2$ TO}
\label{weak_sym}
In this section we give an explanation of the curious dependence of TEN on the entanglement cut, by establishing a connection between the entanglement properties of $\rho_f$ and ground-state $Z_2$ topological order enriched by translation symmetry.
To establish this connection, we first note that at $p=\frac{1}{2}$, $\rho_f$ is nothing but the maximally mixed state with $W_p=A_{p-\bm{\delta}}B_p=1,\forall p$. To find its analog in ground states of local Hamiltonians, it is natural design a stabilizer code with $W_p$ being the stabilizers. Of course, without other terms, we would get a very large ground state degeneracy, and the maximally mixed states in the ground state subspace is just $\rho_f$. Thus we need more stabilizers. Here we provide one illuminating choice:
\begin{equation}
H_{\text{SET}}=-\sum_p W_p-\sum_{\text{vertical link } i}Z_iX_{i+\bm{\delta}}.
\label{eq:SET}
\end{equation}
It is straightforward to check that all terms in the $H_{\text{SET}}$ commute with each other, and the ground state is determined up to topological degeneracy. Actually, this model has been recently constructed and studied in \cite{rao2021theory}. We briefly summarize the important properties of this model: 1. It has $Z_2$ (toric-code) topological order, i.e., it has the same type of anyon excitations and same statistics as the toric code. 2. In this model the $Z_2$ topological order is enriched by translation symmetry along the horizontal direction, which is manifested in the fact that excitations $W_p=-1$ for $p$ on{ even columns} and odd columns belong to different anyon superselection sectors, and correspond to $e$ and $m$ anyons in the toric code, respectively. This phenomenon is often called weak symmetry breaking. 3. As a consequence of weak symmetry breaking, the ground state degeneracy (GSD) on a torus depend on the linear size along the horizontal direction, denoted by $L_x$: GSD$=4 (2)$ for even (odd) $L_x$. 

Our primary goal for introducing this model is to understand the weird behavior of entanglement negativity \eqref{eq:bipartition2} of $\rho_f$ for bipartition 2, so we consider putting the model \eqref{eq:SET} on a cylinder and investigate the entanglement property of the ground state under the same bipartition. First, we note that the ground state, or more specifically, the maximally mixed state in the ground state subspace, can be written in an illuminating way: $\rho_{\text{GS}}\propto\rho_f\prod_{\text{vertical link }i }\frac{1+Z_iX_{i+\bm{\delta}}}{2}$. Moreover, for bipartition 2, the entanglement cut does not go through the stabilizers $Z_iX_{i+\bm{\delta}}$ at all, so these stabilizers contribute zero entanglement. Thus,
\begin{equation}
\varepsilon_A(\rho_{\text{GS}})=\varepsilon_A(\rho_f)=
\left\{
\begin{aligned}
\frac{L_x}{2}\log 2-\log2,\quad  & \text{if } L_x \text{ is even,} \\ 
\frac{L_x}{2}\log 2-\frac{\log2}{2},\quad  & \text{if } L_x \text{ is odd.}
\end{aligned}
\right. 
\label{eq:bipartition2}
\end{equation}
We can instead calculate the entanglement entropy $S_A$ for a (pure) ground state, with a bit more complication. To do this, we need to first specify the boundary condition at the upper and lower boundaries of the cylinder (nevertheless, the result does not depend on the choice of the boundary condition). Then by fixing the value of the logical string operators along the horizontal direction, we can get a pure state and calculate its bipartite entanglement entropy, which yields the same result: 
\begin{equation}
S_A=\left\{
\begin{aligned}
\frac{L_x}{2}\log 2-\log2,\quad  & \text{if } L_x \text{ is even,} \\ 
\frac{L_x}{2}\log 2-\frac{\log2}{2},\quad  & \text{if } L_x \text{ is odd.}
\end{aligned}
\right. 
\end{equation}
Thus, the TEN of $\rho_f$ can be directly related to the TEN/TEE of the ground state of $H_{\text{SET}}$. For the latter, the dependence of TEN/TEE on the parity of $L_x$ is a common feature of topological order with weak symmetry breaking of translations, and can be understood in the following way: Since translations permute $e$ and $m$, for odd $L_x$, $e$ and $m$ are exchanged when going around the cylinder (along the $x$ direction) once. Thus instead of two independent logical string operators along the $x$ direction as naively expected (for even $L_x$, the two logical string operators can be constructed by creating a pair of $e$ or $m$ anyons, dragging one of them around a cycle, and annihilating the pair), one can only find one. This subtlety here causes the TEE/TEN as well as the GSD only half the value as expected for the toric-code topological order.


\subsection{Robustness under phase errors}
\label{rob_app}
In this section, we aim to discuss the robustness of the intrinsic mixed-state topological order under other noises. We demonstrate how to obtain the phase diagram in Fig.~\ref{fig:fig3} with additional single-qubit phase errors. Concretely, we consider the following mixed state:
\begin{equation}
{\rho}_{f,e}=\mathcal{N}^z\circ\mathcal{N}^f[\rho_0].
\end{equation}
Similar to sections A, B, we analyze the property of $\tilde{\rho}_f$ by calculating the von Neumann entropy $S(\tilde{\rho}_f)$ and map it to statistical models. 

We denote the error rate and error chain configuration of $\mathcal{N}^z$ $(\mathcal{N}^f)$ as $p_z$ $(p_f)$ and $C_z$ $(C_f)$, respectively. Then ${\rho}_{f,e}$ can be represented by error chain expansion as in \eqref{eq:errorstring},
\begin{equation}
\rho_{f,e}=\sum_{C_z,C_f}P_f(C_f)P_z(C_z)W^f_{C_f}W^e_{C_z}\rho_0 W^e_{C_z}W^f_{C_f}.
\end{equation}
Then we can write the $n$-th moment as
\begin{equation}
\begin{aligned} 
\Tr(\rho_{f,e}^n)&=\sum_{\left\{C^{(s)}\right\}}\sum_{a^{(s)},b^{(s)}}\prod_{s=1}^{n} \frac{1}{4}P_f\left(C_f^{(s)}\right)P_z\left(C_z^{(s)}\right)\bra{a^{(s)},b^{(s)}}W^e_{C_z^{(s)}}W^f_{C_f^{(s)}} W^f_{C_f^{(s+1)}}W^e_{C_z^{(s+1)}} \ket{a^{(s+1)},b^{(s+1)}}.\\
\end{aligned}
\label{eq:moment2}
\end{equation}
Nonzero contributions only come from error chain configurations satisfying
\begin{equation}
C_z^{(s)}=C_{z}^{(1)}+\partial v^{(s)}+{d_x^{z,(s)}}\gamma_x+{d_x^{z,(s)}}\gamma_y,\ C_{f}^{(s)}=C_{f}^{(1)}+\partial v^{(s)}+{d_x^{f,(s)}}\gamma_x+{d_x^{f,(s)}}\gamma_y,
\end{equation}
so \eqref{eq:moment2} can be simplified as:
\begin{equation}
\begin{aligned}
\Tr(\rho_{f,e}^n)=\frac{1}{4^{n-1}}\cdot\frac{1}{4^{n-1}}\prod_{\alpha=z,f}\left[\sum_{C_\alpha^{(s)}} P\left(C_\alpha^{(1)}\right) \prod_{s=1}^{n-1} \sum_{\left\{v_\alpha^{(s)}\right\}}\sum_{d_x^{\alpha,(s)},d_y^{\alpha,(s)}=0,1} P\left(C_\alpha^{(1)}+\partial v_\alpha^{(s)}+d_x^{\alpha,(s)}\gamma_x+{d_y^{\alpha,(s)}}\gamma_y\right)\right].
\end{aligned}
\end{equation}
As in \eqref{eq:RBIM2}, the collection of terms in the bracket for each $\alpha$ can be mapped to the partition function of a $(n-1)$-flavor RBIM: 

\begin{equation}
\begin{aligned}
\Tr(\rho_{f,e}^n)=\frac{1}{4^{2n-2}}\left(\sqrt{(1-p_f)p_f}\right)^{(n-1)N}\left(\sqrt{(1-p_z)p_z}\right)^{(n-1)N}\overline{\left(Z^{\text{RBIM}}_{p_f}(J_f)\right)^{n-1}}\cdot\overline{\left(Z^{\text{RBIM}}_{p_z}(J_z)\right)^{n-1}},
\end{aligned}
\label{eq:RBIM3}
\end{equation}
where $J_\alpha$ $(\alpha=z,f)$ is the strength of Ising coupling for each of the two RBIMs. $p_\alpha$ denotes the probability of antiferromagnetic coupling on each bond. Both RBIMs are situated along the Nishimori line: $e^{-2J_\alpha}=\frac{p_\alpha}{1-p_\alpha}$. Again, the partition functions implicitly contain summations over the four boundary conditions. The von Neumann entropy can be obtained by taking the $n\rightarrow 1$ limit:
\begin{equation}
\begin{aligned}
S({\rho}_{f,e})&=- \lim_{n\rightarrow1}\frac{\partial }{\partial n}\Tr({\rho}_{f,e})\\
&=-\overline{ \log Z^{\text{RBIM}}_{p_z}(J_z)}-\overline{ \log Z^{\text{RBIM}}_{p_f}(J_f)} +\left(-4\log 2-\frac{N}{2}\log[p_z(1-p_z)]-\frac{N}{2}\log[p_f(1-p_f)]\right),
\end{aligned}
\end{equation}
The terms in the parentheses is always regular for finite $p_{z},p_f$, so we can focus on the first two terms, which is the free energy of two decoupled RBIMs. We denote the Ising spin variables of the two RBIMs as $\sigma$ and $\tau$, respectively. Then we can straightforwardly get the phase diagram of the statistical model, shown in Fig.~\ref{fig:fig3}. 



Applying the same strategy as in sections A, B, we can map the relative entropy and coherent information to observables in the RBIM. We directly list the results here. 
\begin{enumerate}
\item Relative entropy. $D(\rho_{f,e}||\rho^e_{f,e})$ is mapped to the excess free energy of the defect line (connecting the inserted pair of $e$ anyons) of the RBIM of spin $\sigma$; $D(\rho_{f,e}||\rho^f_{f,e})$ is mapped to the excess free energy of the defect line (connecting the inserted pair of $f$ anyons) of the RBIM of spin $\tau$;  $D(\rho_{f,e}||\rho^m_{f,e})$  is mapped to the sum of the excess free energy of the defect line (connecting the inserted pair of $m$ anyons) of the two RBIMs, because $w^m=w^ew^{f}$.
\item Coherent information,
\begin{equation}
    I_c=2\log 2-\overline{\log\left[\sum_{d_x,d_y=0,1 }e^{-\Delta F^\sigma_{d_x,d_y}}\right]}-\overline{\log\left[\sum_{d_x,d_y=0,1 }e^{-\Delta F^\tau_{d_x,d_y}}\right]},
\end{equation}

\end{enumerate}

where $\Delta F_{d_x,d_y}^{\sigma(\tau)}$ is the excess free energy of non-contractible defect lines in the RBIM of spin $\sigma\ (\tau)$. Across transitions at $p_z\approx 0.109$ and $p_f\approx 0.109$, $I_c$ changes discontinuously. Quantum memory can only be realized when both $\sigma$ and $\tau$ are in the FM phase, while classical memory corresponds to one of the spin species is in the FM phase while the other in the PM phase, and the topological memory is completely lost when both RBIMs are in the PM phase.

From the above mapping, we can relate the four phases of the RBIM to the four types of topological order (including the trivial one) in Fig.~\ref{fig:fig3} . 


\section{Exact solution to the gapless spin liquid phase of the toric code model through fermionization }
\label{app:fermionization}
\begin{figure}[htb]
 \centering
\includegraphics[width=0.45\linewidth]{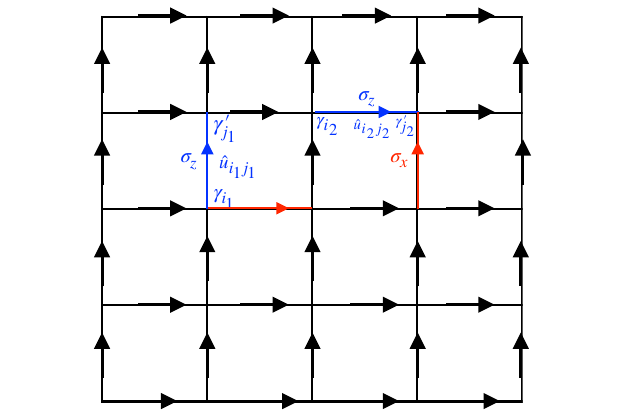}

\caption{The fermionization of the model in \eqref{eq:ZX}. The links that Pauli matices $Z_i, X_{i+\bm{\delta}}$ act upon are colored in blue and red, respectively. There are two Majorana degrees of freedom $\gamma_{v_i},\gamma^{\prime}_{v_i}$ on each lattice site $v_i$ in the fermionized Hilbert space, corresponding to the $f$ anyon in the toric code model and we implicitly assume that the location of the $f$ anyon is the same as the comprising $e$ anyon. The term $Z_i X_{i+\bm{\delta}}$ are fermionized as  $iu_{i}\gamma_{v_i}\gamma'_{v'_i}$, where $v_i,v'_i$ is the starting point and the end point of the link $i$, with the direction defined by the arrows, and $u_i$ is a static $Z_2$ gauge field on link $i$, which accounts for the mutual semion statistics between $f$-anyons (fermions) and  $m$-anyons ($Z_2$ flux).  }

\label{fig:fermionization}
\end{figure}
In this section we analyze the properties the toric code model with additional $ZX$ terms in the Hamiltonian,
\begin{equation}
H=-\sum_v A_v-\sum_p B_p -\sum_i h_{xz} Z_i X_{i+\bm{\delta}}.
\label{eq:ZX}
\end{equation}
We show that this model can be solved exactly using the method introduced in \cite{chen2018bosonization}. Firstly, we note that the model has an extensive number of locally conserved quantities. $[H,W_p]=0$ with $W_p=A_{v=p-\bm{\delta}} B_{p}$, which follows from the fact that the $e$ anyons and adjacent $m$ anyons are always created or annihilated in pairs, so we can solve the model in each simultaneous eigenspace of $W_p$. Secondly, the role of $ZX$ term is to induce pair creation, annihilation and hopping of $f$ anyons, which are fermions. Then on an infinite lattice or a topologically trivial lattice, in each sector $\{W_p=w_p\}$, the only degrees of freedom are the $f$ anyons, so we expect in each sector the model can be described by a fermion tight-binding model. We assume the fermions are defined on the vertices of the lattice, with the mapping 
\begin{equation}
n^f_v\longleftrightarrow \frac{1-A_v}{2}, 
\label{eq:map1}
\end{equation}
where $n^f_v=f^\dagger_v f_v$ is the fermion number operator. This mapping follows naturally from the observation that in the zero-flux sector $\{w_p=1\}$, $\frac{1-A_v}{2}$ corresponds to the occupation number of the $f$ anyon on $v$.  Finally, because $f$ and $e/m$ anyons are mutual semions, an $f$ anyon can acquire a nontrivial phase depending on $w_p$ when moving around the plaquette $p$. Thus $W_p$ should correspond to static $Z_2$ flux on each plaquette in the fermion model, so we have the following mapping,
\begin{equation}
T_{v_iv'_i}\equiv iu_{i}\gamma_{v_i}\gamma'_{v'_i} \longleftrightarrow Z_i X_{i+\bm{\delta}} ,\quad \text{link }i\equiv \langle v_iv'_i\rangle
\label{eq:map2}
\end{equation}
where $\gamma_v=f_v+f^\dagger_v,\gamma'_{v'}=-i(f_{v'}-f^\dagger_{v'})$ are Majorana fermion operators and $u_{i}=\pm 1$ are static $Z_2$ gauge fields defined on links. It is straightforward to check that the commutation and anti-commutation relation between $Z_i X_{i+\bm{\delta}}$ is preserved under the above mapping:
\begin{equation}
\left\{
\begin{aligned}
\{T_{v_iv'_i},T_{v_jv'_j}\}=0,\quad  & \text{if } i=j\pm \bm{\delta}\ \\ 
{[T_{v_iv'_i},T_{v_jv'_j}]=0,}\quad & \text { otherwise }
\end{aligned}
\right. \longleftrightarrow 
\left\{\begin{array}{cl}
\{Z_iX_{i+\bm{\delta}},Z_jX_{j+\bm{\delta}}\}=0,\quad & \text{if } i=j\pm \bm{\delta} \\
{[Z_iX_{i+\bm{\delta}},Z_jX_{j+\bm{\delta}}]=0,}\quad & \text{ otherwise } 
\end{array}
\right.
\end{equation}

 The commutation and anti-commutation relation between $Z_iX_{i+\bm{\delta}}$ and $A_v$ (and similarly for $B_p=A_{v=p-\bm{\delta}}W_p$) is also preserved:
 \begin{equation}
\left\{
\begin{aligned}
\{T_{v_iv'_i},1-2n^f_v\}=0,\quad  & \text{if } v\in\partial i \\ 
{[T_{v_iv'_i},1-2n^f_v]=0,}\quad & \text { otherwise }
\end{aligned}
\right. \longleftrightarrow 
\left\{\begin{array}{cl}
\{Z_iX_{i+\bm{\delta}},A_v\}=0,\quad &\text{if } v\in\partial i \\
{[Z_iX_{i+\bm{\delta}},A_v]=0, }\quad &\text{otherwise } 
\end{array}
\right.
\end{equation}
 Besides, $Z_iX_{i+\bm{\delta}}$ and $A_v,B_p$ satisfy an additional relation: 
 \begin{equation}
 \prod_{i\in\partial p}Z_iX_{i+\bm{\delta}}=B_p A_{v=p+\bm{\delta}}
 \label{eq:additional}.
 \end{equation}
 
 Under the mapping in \eqref{eq:map2}, the left hand side of \eqref{eq:additional} is mapped to $\prod_{i\in\partial p}T_{v_iv'_i}=(1-2n^f_{p-\bm{\delta}})(1-2n^f_{p+\bm{\delta}})\prod_{i\in\partial p}\hat{u}_{i}$. The right hand side of  \eqref{eq:additional} can be rewritten as $W_pA_{p-\bm{\delta}}A_{p+\bm{\delta}}$. Then \eqref{eq:additional} together with the \eqref{eq:map1}) determines the $Z_2$ flux configuration in the fermion model, 
 \begin{equation}
 \prod_{i\in\partial p}\hat{u}_{i}\longleftrightarrow A_{p-\bm{\delta}}B_{p}=W_p
 \label{eq:map3},
 \end{equation}
as expected.
\eqref{eq:map1},\eqref{eq:map2},\eqref{eq:map3} form the complete the dictionary of the fermionization procedure on an infinite lattice or a topologically trivial lattice. However, on the fermionic side, under periodic boundary condition, i.e., on a torus, there are additional $Z_2$ fluxes threading the two non-contractible cycles $\gamma_x,\gamma_y$ along the $x,y$ direction: $\hat{w}_{x,y}=\prod_{i\in\gamma_{x,y}}\hat{u}_i$. We need to figure out what is the counterpart of $\hat{w}_{x,y}$ on the toric code side. This can be done by using again the mapping in \eqref{eq:map2}, which leads to
\begin{equation}
-\left(\prod_{i\in\gamma_{x,y}}\hat{u}_i\right)\prod_{i\in\gamma_{x,y}}(1-2n^f_{v_i}) \longleftrightarrow \prod_{i\in\gamma_{x,y}}Z_iX_{i+\bm{\delta}}.
\end{equation}
By using \eqref{eq:map1}, we obtain,
\begin{equation}
-\hat{w}_{x(y)}\longleftrightarrow \prod_{i\in\gamma_{x(y)}}A_{v_i}Z_iX_{i+\bm{\delta}}=\prod_{i\in\gamma_{x(y)}}Z_iX_{i-\bm{\delta}}\equiv\hat{W}^{f'}_{\gamma_{x(y)}}.
\end{equation}
Indeed, $\hat{W}^{f'}_{\gamma_x},\hat{W}^{f'}_{\gamma_y}$ are also conserved quantities in the original model, $[\hat{W}^{f'}_{\gamma_{x(y)}},H]=0$. 

In the end, we map the model in \eqref{eq:ZX} to a quadratic fermion model with static $Z_2$ gauge field,
\begin{equation}
H\leftrightarrow \tilde{H}=\sum_v(2n^f_v-1)\cdot(1+\hat{w}_p)-h_{xz}\sum_{\langle vv'\rangle}i\hat{u}_{vv'}\gamma_v\gamma'_{v'}, 
\end{equation}
where $\hat{w}_p=\prod_{\langle vv'\rangle\in \partial p}\hat{u}_{vv'}$. 

Thanks to the extensive number of conserved quantities, $[\hat{u}_{vv'},\tilde{H}]=[\hat{w}_{p/x/y},\tilde{H}]=0$, $\tilde{H}$ can be reduced to a free fermion model in each $Z_2$ flux sector $\{\hat{w}_p=w_p,\hat{w}_x=w_x,\hat{w}_y=w_y\}$, and thus can be easily solved. In the case $h_{xz}=0$, it is obvious that the ground state (the vaccum of $f$) lies in the zero flux sector $w_p=1$, and the lowest energy state in the four sectors with distinct $\{w_x=\pm 1,w_y=\pm 1\}$ has degenerate eigenenergy. This is just another viewpoint of the well-known topological degeneracy. 

Via numerical investigation we find that the ground state always stay in the sector with $w_p=1$, irrespective of the value of $h_{xz}$, so we will mainly restrict our discussion to this case,
\begin{equation}
    \tilde{H}=4\sum_v n^f_v-h_{xz}\sum_{\langle vv'\rangle} f^\dagger_vf_{v'}+f_vf_{v'}+h.c.,
\end{equation}
and ${w_{a}}=1,-1(a=x,y)$ corresponds to PBC and APBC along direction $a$, respectively. Then $\tilde{H}$ can be solved via Fourier transformation, $f_v=\frac{1}{\sqrt{L_xL_y}}\sum_{k_a=\frac{2n_a\pi}{L_a}} f_k e^{ik_x{v_x} +ik_yv_y}$, where $n\in \mathbb{Z}$ for PBC and $n\in \mathbb{Z}+\frac{1}{2}$ for APBC,

\begin{equation}
\tilde{H}=\sum_k (f^\dagger_k, f_{-k})\left(\begin{array}{cc}
	2-h_{xz}(\cos k_x+\cos k_y) & -ih_{xz}(\sin k_x+\sin k_y) \\
	ih_{xz}(\sin k_x+\sin k_y) & -2+h_{xz}(\cos k_x+\cos k_y)
\end{array} \right)\left(\begin{array}{c}
	f_k \\
	f^\dagger_{-k}
\end{array} \right).
\end{equation}

The dispersion of Bogoliubov quasiparticle excitation can be easily obtained,
\begin{equation}
\xi_k=2\sqrt{[2-h_{xz}(\cos k_x+\cos k_y)]^2+[h_{xz}(\sin k_x+\sin k_y)]^2},
\end{equation}
and the ground state energy is $E_g=-\sum_k\frac{\xi_k}{2}$. For $h_{xz}<1$, the spectrum is gapped and the ground energy is nearly degenerate (with an exponentially small splitting) for the 4 types of boundary conditions. This corresponds to the gapped topologically ordered phase of $H$. For $h_{xz}=1$, the gap closes at $k_x=k_y=0$, and remain closed for $h_{xz}>1$, with linear dispersion at two Dirac points $k_x=-k_y=\pm \arccos \frac{1}{h_{xz}}$. So for $h_{xz}>1$ the original model lies in a gapless spin liquid phase, reminiscent of the gapless phase of the Kitaev honeycomb model. In this phase the topological degeneracy is lifted by an algebraically small gap, but the ground state remains long-range entangle.

\section{Details about the decohered Kitaev honeycomb model}
\subsection{The model}
In this Appendix, we provide detailed analysis of the effect proliferation of the $f$ anyons in the Kitaev honeycomb model at zero magnetic field:
\begin{equation}
H=-J_{x} \sum_{x \text {-bonds }} \sigma_{j}^{x} \sigma_{k}^{x}-J_{y} \sum_{y \text {-bonds }} \sigma_{j}^{y} \sigma_{k}^{y}-J_{z} \sum_{z \text {-bonds }} \sigma_{j}^{z} \sigma_{k}^{z}.
\label{kitaev_ham}
\end{equation}
It can be exactly solved by introducing the Majorana fermion operators: ${\sigma^\alpha}=i{b^\alpha}b^0$. After fixing the $Z_2$ gauge fields as $\hat{u}_{j k}=i b_{j}^{\alpha} b_{k}^{\alpha}=1$ (which corresponds to the zero gauge flux sector the ground state lies in), the Kitaev Hamiltonian \eqref{kitaev_ham} becomes the following quadratic fermion model:
\begin{equation}
H_F=-J_x\sum_{x \text {-bonds }} b_{j}^{x} b_{k}^{x}-J_{y} \sum_{y \text {-bonds }} b_{j}^{y} b_{k}^{y}-J_{z} \sum_{z \text {-bonds }} b_{j}^{z} b_{k}^{z},
\end{equation}
whose ground state $|\psi_F\rangle$ can be easily solved. The physical ground state $|\Psi\rangle$ can be obtained by projection to the gauge invariant subspace using the projection operator $\hat{P}=\Pi_{i}\frac{1+\hat{D}_i}{2}$, i.e., $|\Psi\rangle=\Pi_{i}\frac{1+\hat{D}_i}{2}|\{u_{ij}=1\}\rangle\otimes|\psi_F\rangle$, where $\hat{D}_i=b^{x}_i b^{y}_i b^{z}_i b^0_i$. We will focus on Abelian phase with $|J_z|>|J_x|+|J_y|$, where $H_F$ is trivially gapped and $|\Psi\rangle$ belongs to the $\Z_2$ TO.

The effect of the noisy channel in \eqref{decoheredhoneycomb} is to change the fermion part of the density matrix but leave the gauge field configuration invariant. This can be shown by writing the Kraus operators in terms of Majorana fermion operators: $\sigma^{\alpha}_{i}\sigma^{\alpha}_{j}=-i\hat{u}_{ij}b_i^0b_j^0$. Thus the density matrix $\rho_f$ can be written as: 
\begin{equation}
    \rho_f=\hat{P}|\{u_{ij}=1\}\rangle\langle\{u_{ij}=1\}|\otimes\rho_F\hat{P}, 
\end{equation}
where $\rho_F$ is the Majorana fermion density matrix, which is $\rho_F=\mathcal{N}^{X,F} \circ \mathcal{N}^{Y,F}\circ \mathcal{N}^{Z,F}\left[\rho_{F,0}\right]$, where $\rho_{F,0}=|\psi_F\rangle\langle\psi_F|$ is the ground state of $H_F$. $\mathcal{N}^{\alpha,F}=\prod_{\langle ij \rangle} \mathcal{N}^{\alpha,F}_{\langle ij \rangle}$, with
\begin{equation}  
\mathcal{N}^{\alpha,F}_{\langle ij\rangle\in \alpha \text {-bonds }}[\cdot] \equiv\left(1-p\right) \cdot+p b_i^0b_j^0\cdot b_j^0b_i^0.
\end{equation}
Similar to the case in the toric code model, we expect that the error-corrupted state $\rho_f$ undergoes a phase transition in the mixed-state topological order at some critical error rate $p_c$. In the limit $|J_z|\gg |J_x|,|J_y|$, $p_c$ can be determined by mapping to the RBIM, analogous to the case in the toric code, which gives $p_c\approx 0.109$. The topological quantum memory also breaks down to classical topological memory after the transition, with two remaining commuting logical operators:
\begin{equation}
W_{\gamma_x}=\prod_{\langle ij\rangle\in\gamma_x}\sigma_i^{\alpha_{\langle ij\rangle}}\sigma_j^{\alpha_{\langle ij\rangle}},\ W_{\gamma_y}=\prod_{\langle ij\rangle\in\gamma_y}\sigma_i^{\alpha_{\langle ij\rangle}}\sigma_j^{\alpha_{\langle ij\rangle}},
\end{equation}
where $\alpha_{\langle ij\rangle}=x,y,z$  for $\langle ij\rangle \in$ $x$-bonds, $y$-bonds, $z$-bonds. The two logical operators are depicted in Fig.~\ref{fig:mto_kitaev}. The residual classical memory is due to the fact that $[W_{\gamma_x},\sigma_i^{\alpha_{\langle ij\rangle}}\sigma_j^{\alpha_{\langle ij\rangle}}]=[W_{\gamma_y},\sigma_i^{\alpha_{\langle ij\rangle}}\sigma_j^{\alpha_{\langle ij\rangle}}]=0, \ \forall \langle ij \rangle.$

\subsection{Entanglement negativity}
In this section, we compute the entanglement negativity of the $\rho_f$ in the decohered Kitaev honeycomb model, starting from the Abelian phase. We will demonstrate the existence of nonzero TEN for any error rate $p$. 

Again, we use the replica trick to compute the entanglement negativity:
\begin{equation}
\mathcal{E}_{A}(\rho_f):=\log \left\|\rho_f^{T_{A}}\right\|_{1}=\lim_{2n\rightarrow 1}\frac{1}{2-2 n} \log \frac{\Tr\left(\rho_f^{T_{A}}\right)^{2 n}}{\Tr \rho_f^{2 n}}.
\end{equation}
First it is easy to show: $\lim_{2n\rightarrow 1} \Tr \rho_{f}^{2 n}=\Tr \rho_{f}=1$, due to the trace-preserving property of quantum channels. So we only need to deal with the numerator: $\Tr\left(\rho_f^{T_{A}}\right)^{2 n}$.

A general curve $\gamma$, which bipartites the honeycomb lattice into subregions $A$ and $B=\bar A$, intersects the bonds of the honeycomb lattice. In order to partial transpose the degrees of freedom in $A$ subregion, we should define new $Z_2$ gauge fields to replace the gauge fields on the intersected bonds. Following the notation in \cite{PhysRevLett.105.080501}, we assume $\gamma$ intersects $2L$ bonds, and we denote the bonds intersected by the curve $\gamma$ as: $\langle a_n b_n\rangle,n=1,2,...,2L$. If the $Z_2$ gauge field on the bonds $\langle a_{2n-1}b_{2n-1}\rangle$ and $\langle a_{2n}b_{2n}\rangle$  are $\hat{u}_{a_{2n-1} b_{2n-1}}=ib_{a_{2n-1}}^\alpha b_{b_{2n-1}}^\alpha,\hat{u}_{a_{2n} b_{2n}}=ib_{a_{2n}}^\beta b_{b_{2n}}^\beta $; then we introduce two $Z_2$ gauge fields in the A and B subregions respectively: $w_{A,n}=ib^{\alpha}_{a_{2n-1}}b^{\beta}_{a_{2n}},w_{B,n}=ib^{\alpha}_{b_{2n-1}}b^{\beta}_{b_{2n}}$. Then the ground state configuration of the gauge fields on the intersected links can be written as:
\begin{equation}
\left|\{u_{p}\}\right\rangle=\frac{1}{\sqrt{2^{L}}} \sum_{w_{A}=w_{B}=\{ \pm 1\}}\left|w_{A}, w_{B}\right\rangle,
\end{equation}
where $\ket{\{u_p\}}$ is the direct product of  $\ket{u_{a_n b_n}=1}$. As a result, the ground state density matrix can be written as:
\begin{equation}
\begin{aligned}
    \rho_0&=\frac{1}{2^{N+L+1}} \sum_{g,g^{\prime}, w,w^{\prime}} D_{g}\ket{u_{A}, w} \ket{u_{B},w}\bra{u_A,w^{\prime}}\bra{u_B,w^{\prime}}\otimes \rho_{F,0}D_{g^{\prime}},
\end{aligned}
\end{equation}
where the summation over $g,g^{\prime}$ is over all the possible sets of the lattice sites, and $D_g=\prod_{i\in g} D_i$.  What's more, we can simply replace the $\rho_{F,0}$ with $\rho_{F}$ to get the decohered density matrix $\rho_f$. The partial trace of the density matrix is:
\begin{equation}
    \rho_f^{T_{A}}=\frac{1}{2^{N+L+1}} \sum_{g,g^{\prime}, w,w^{\prime}} D_{g_A^{\prime}}D_{g_B}\ket{u_{A}, w^{\prime}} \ket{u_{B},w}\bra{u_A,w}\bra{u_B,w^{\prime}}\otimes \rho^{T_A}_F D_{g_B^{\prime}}D_{g_A}.
\end{equation}
Thus
\begin{equation}
    \begin{aligned}
        (\rho_f^{T_{A}})^2=&(\frac{1}{2^{N+L+1}})^2 \sum_{g,g^{\prime}, w,w^{\prime}}\sum_{g_2,g_2^{\prime}, w_2,w_2^{\prime}} \bra{u_A,w}\bra{u_B,w^{\prime}} D_{g^{\prime}_B}D_{g_A}D_{g_{2,B}}D_{g^{\prime}_{2,A}}\ket{u_{A}, w_2^{\prime}}\ket{u_{B},w_2}\\
        &D_{g_A^{\prime}}D_{g_B}\ket{u_{A}, w^{\prime}}\ket{u_{B},w} \bra{u_{A}, w_2}\bra{u_{B},w_2^{\prime}}D_{g'_{2B}}D_{g'_{2A}}\otimes (\rho_F^{T_A})^2
    \end{aligned}
\end{equation}
where $g_A$ ,$g_B$, are the sets of lattice sites $g\bigcap A, g\bigcap B$, respectively. Now we split $D_g$ into the gauge field part and fermion part: $D_g=X_gY_g$, where $X_{g}=i^{|g|(|g|-1) / 2} \prod_{j \in g} b_{j}^{x} b_{j}^{y} b_{j}^{z}$ and $Y_{g}=i^{|g|(|g|-1) / 2} \prod_{j \in g} b^0_{j}$, where $|g|$ is the number of lattice sites in region $g$. The inner product can be simplified as:
\begin{equation}
\begin{aligned}
       & \bra{u_A,w}\bra{u_B,w^{\prime}} D_{g_B^{\prime}}D_{g_A}D_{g_{2,B}}D_{g_{2,A^{\prime}}}\ket{u_{A}, w_2^{\prime}}\ket{u_{B},w_2}\\
       &=\delta_{w, w_2^{\prime}}\left(\delta_{g_{A}, g_{2,A}^{\prime}}+x_{A}(w) \delta_{g_{A}, A-g^{\prime}_{2,A}}Y_{A}\right)\delta_{w^{\prime}, w_2}\left(\delta_{g^{\prime}_{B}, g_{2,B}}+x_{A}(w) \delta_{g^{\prime}_{B}, B-g_{2,B}}Y_{B}\right)\\
       &=\left(2\delta_{w, w_2^{\prime}}P_{F,A}^{x_A(w)}\right)\left(2\delta_{w^{\prime}, w_2}P_{F,B}^{x_B(w^{\prime})}\right),
\end{aligned}
\label{eq:innerproduct}
\end{equation}
where $P_{F,A(B)}^{x}=\frac{1+xY_{A(B)}}{2}$ is the projection to the subspace with fixed Fermi parity $x$ in subregion $A$ $(B)$, and $x_{A(B)}(w)=\left\langle u_{A(B)}, w\left|X_{A(B)}\right| u_{A(B)}, w\right\rangle=p_{A(B)} \prod_{n=1}^{L} w_{n}$, where we define $p_{A(B)}\equiv\prod_{i,j \in A(B)} u_{i j}$.  Puting this inner product back into the $(\rho^{T_{A}})^2$, we obtain:
\begin{equation}
    \begin{aligned}
        (\rho_f^{T_{A}})^2&=(\frac{1}{2^{N+L+1}})^2 \sum_{g,g^{\prime}, w,w^{\prime}} D_{g}\ket{u_{A}, w^{\prime}}\ket{u_{B},w}\bra{u_{A}, w^{\prime}}\bra{u_{B},w} \otimes \rho_F^{T_A}2^N\left(2P_{F,A}^{x_A(w)}\right)\left(2P_{F,B}^{x_B(w^{\prime})}\right)\rho_F^{T_A}D_{g^{\prime}}\\
        &=\frac{1}{2^{N+2L}} \sum_{g,g^{\prime}, w,w^{\prime}} D_{g}\ket{u_{A}, w^{\prime}}\ket{u_{B},w}\bra{u_{A}, w^{\prime}}\bra{u_{B},w} \otimes \rho_F^{T_A}P_{F,A}^{x_A(w)}P_{F,B}^{x_B(w^{\prime})}\rho_F^{T_A}D_{g^{\prime}}
    \end{aligned}
\end{equation}
Now we move one step further to calculate $(\rho^{T_{A}})^4$:
\begin{equation}
\begin{aligned}
        (\rho_f^{T_{A}})^4&= \left((\rho^{T_{A}})^2\right)^2=(\frac{1}{2^{N+2L}})^2 \sum_{g,g^{\prime}, w,w^{\prime}} D_{g}\ket{u_{A}, w^{\prime}}\ket{u_{B},w}\bra{u_{A}, w^{\prime}}\bra{u_{B},w} \otimes \rho_F^{T_A}P_{F,A}^{x_A(w)}P_{F,B}^{x_B(w^{\prime})}\rho_F^{T_A}D_{g^{\prime}}\\
        &\sum_{g_2,g_2^{\prime}, w_2,w_2^{\prime}} D_{g_2}\ket{u_{A}, w_2^{\prime}}\ket{u_{B},w_2}\bra{u_{A}, w_2^{\prime}}\bra{u_{B},w_2} \otimes \rho_F^{T_A}P_{F,A}^{x_A(w_2)}P_{F,B}^{x_B(w_2^{\prime})}\rho_F^{T_A}D_{g_2^{\prime}}\\
        &=(\frac{1}{2^{N+2L}})^2 \sum_{g,g^{\prime}, w,w^{\prime}} D_{g}\ket{u_{A}, w^{\prime}}\ket{u_{B},w}\bra{u_{A}, w^{\prime}}\bra{u_{B},w}\otimes \rho_F^{T_A}P_{F,A}^{x_A(w)}P_{F,B}^{x_B(w^{\prime})}\rho_F^{T_A}\\
        &2^N\left(2P_{F,A}^{x_A(w^{\prime})}\right)\left(2P_{F,B}^{x_B(w)}\right)\rho_F^{T_A}P_{F,A}^{x_A(w)}P_{F,B}^{x_B(w^{\prime})}\rho_F^{T_A}D_{g_2^{\prime}}\\
        &=\frac{1}{2^{N+4L-2}}\sum_{g,g^{\prime}, w,w^{\prime}} D_{g}\ket{u_{A}, w^{\prime}}\ket{u_{B},w}\bra{u_{A}, w^{\prime}}\bra{u_{B},w}\\
        &\otimes \left(\rho_F^{T_A}P_{F,A}^{x_A(w)}P_{F,B}^{x_B(w^{\prime})}\right)\left(\rho_F^{T_A}P_{F,A}^{x_A(w^{\prime})}P_{F,B}^{x_B(w)}\right)\left(\rho_F^{T_A}P_{F,A}^{x_A(w)}P_{F,B}^{x_B(w^{\prime})}\right)\rho_F^{T_A} D_{g^{\prime}}.
    \end{aligned}
\end{equation}
With these results, we can now arrive at $(\rho^{T_{A}})^{2n}$ by the iteration and induction:
\begin{equation}
\begin{aligned}
    (\rho_f^{T_{A}})^{2n}&=\frac{1}{2^{N+2nL-2(n-1)}}\sum_{g,g^{\prime}, w,w^{\prime}} D_{g}\ket{u_{A}, w^{\prime}}\ket{u_{B},w}\bra{u_{A}, w^{\prime}}\bra{u_{B},w}\\
    &\left[\left(\rho_F^{T_A}P_{F,A}^{x_A(w)}P_{F,B}^{x_B(w^{\prime})}\right)\left(\rho_F^{T_A}P_{F,A}^{x_A(w^{\prime})}P_{F,B}^{x_B(w)}\right)\right]^{n-1}\left(\rho_F^{T_A}P_{F,A}^{x_A(w)}P_{F,B}^{x_B(w^{\prime})}\right)D_{g^{\prime}}.
    \end{aligned}
\end{equation}
Using \eqref{eq:innerproduct} again, we can obtain the trace:
\begin{equation}
\begin{aligned}
    \Tr(\rho_f^{T_{A}})^{2n}&=\frac{1}{2^{2nL-2n}}\sum_{w,w^{\prime}} \Tr_F\left(\rho_F^{T_A}P_{F,A}^{x_A(w^{\prime})}P_{F,B}^{x_B(w)}\rho_F^{T_A}P_{F,A}^{x_A(w)}P_{F,B}^{x_B(w^\prime)}\right)^n\\
    &=\frac{2^{L-1}\cdot2^{L-1}}{2^{2n(L-1)}}\Tr_F\bigg(\rho_F^{T_A}P_{F,A}^{p_A}P_{F,B}^{p_B}\rho_F^{T_A}P_{F,A}^{p_A}P_{F,B}^{p_B}+\rho_F^{T_A}P_{F,A}^{p_A}P_{F,B}^{-p_B}\rho_F^{T_A}P_{F,A}^{-p_A}P_{F,B}^{p_B}\\
    &\qquad\qquad\qquad\qquad+\rho_F^{T_A}P_{F,A}^{-p_A}P_{F,B}^{p_B}\rho_F^{T_A}P_{F,A}^{p_A}P_{F,B}^{-p_B}+\rho_F^{T_A}P_{F,A}^{-p_A}P_{F,B}^{-p_B}\rho_F^{T_A}P_{F,A}^{-p_A}P_{F,B}^{-p_B}\bigg)^n
    \end{aligned}
\end{equation}
Since the projector $D_{\text{tot}}=\prod_i D_i=X_{\text{tot}}Y_{\text{tot}}=1$, the total fermion parity of the whole system is fixed by $Y_{\text{tot}}=X_{\text{tot}}=p_Ap_B$. Therefore, terms in the bracket can be simplified as:
\begin{equation}
\begin{aligned}
&\rho_F^{T_A}P_{F,A}^{p_A}P_{F,B}^{p_B}\rho_F^{T_A}P_{F,A}^{p_A}P_{F,B}^{p_B}+\rho_F^{T_A}P_{F,A}^{p_A}P_{F,B}^{-p_B}\rho_F^{T_A}P_{F,A}^{-p_A}P_{F,B}^{p_B}+\rho_F^{T_A}P_{F,A}^{-p_A}P_{F,B}^{p_B}\rho_F^{T_A}P_{F,A}^{p_A}P_{F,B}^{-p_B}+\rho_F^{T_A}P_{F,A}^{-p_A}P_{F,B}^{-p_B}\rho_F^{T_A}P_{F,A}^{-p_A}P_{F,B}^{-p_B}\\
=&\rho_F^{T_A}(P^+_{F,A}+P^-_{F,A})(P^+_{F,B}+P^-_{F,B})\rho_F^{T_A}(P^+_{F,A}+P^-_{F,A})(P^+_{F,B}+P^-_{F,B})\\
=&(\rho^{T_A}_{F})^2
\end{aligned}
\end{equation}
Then we can finally get the entanglement negativity:
\begin{equation}
    \begin{aligned}
        \mathcal{E}_{A}(\rho_f)&=\log(\tr(\rho_f^{T_A}))\\
        &=\lim_{2n\rightarrow 1}\frac{1}{2-2 n} \log \tr\left(\rho^{T_{A}}\right)^{2 n}\\
        &=L\log 2-\log 2 +\log||\rho_F^{T_A}||_1,
    \end{aligned}
\end{equation}
where the last term is the entanglement negativity $\varepsilon_A(\rho_F)$ of the density matrix of fermions. We note that the above result holds for any parameters $J_x,J_y,J_z$, and also for general bipartitions as long as the length of the boundary is even ($N_l=2L$) and $A,B$ are connected. To analyze the scaling behavior of $\varepsilon_A$, we again consider bipartition of a cylinder with translation invariant entanglement cut. In the gapped phase, $\rho_{F,0}$ is a Gaussian state with finite correlation length, so $\varepsilon_A(\rho_{F,0})= \alpha L+\cdots$ satisfies an area law with a vanishing subleading term for $L\rightarrow \infty$, 
We expect $\varepsilon_A(
\rho_{F})$ also have the same property after applying local channels on $\rho_F$, since local quantum channels cannot generate long-range entanglement. Thus we get TEN $=\log2$, consistent with our expectation for even boundary sizes, which shows that this result holds for mixed-state topological order beyond stabilizer codes. For the decohered Kitaev honeycomb model, we still anticipate TEN to depend on the parity of $N_l$, like in the toric code model. Indeed, in the case $p=\frac{1}{2}$, $\rho_f$ is equivalent to $\rho_f$ up to an onsite unitary transformation. However, for general $p$, the calculation of negativity for odd $N_l$ is more complicated and thus not shown here.    

\section{A proof that $e^2m^2$ is a deconfined anyon\label{app:deconfinement}}
In this Appendix we prove that $e^2m^2$ indeed becomes a deconfined anyon in the decohered double semion model, according to Definition 1. To start with, we write down the explicit form of the decohered state $\rho=\mathcal{N}^{[em]}[\rho_0]$, with maximal decoherence $p_0=p_1=p_2=p_3=\frac{1}{4}$. As discussed below \eqref{eq:DS_codespace}, $\rho$ is determined by the stabilizer group $G=\langle \{W^{e^3m}_p\},\text{nonlocal stabilizers}\rangle$,
\begin{equation}
\rho=\frac{1}{4^N}\sum_{g\in G}g.
\end{equation}

We define the open $e^2m^2$ string $W^{e^2m^2}_{\tilde{C}}=\prod_{i\in\tilde{C}}X^2_iZ^2_{i+\bm{\delta}}$ on the dual lattice and $U_{\tilde{C}}=\frac{I+iW^{e^2m^2}_{\tilde{C}}}{\sqrt{2}}$. Then $e^2m^2$ anyons can be created at $\partial\tilde{C}$ by $\rho\rightarrow U_{\tilde{C}}\rho U^\dagger_{\tilde{C}}$. We prove below that both criteria in the definition of deconfined excitations are satisfied. The second criterion directly follows from the fact that $e^2m^2$ anyons generate a strong $1$-form symmetry of $\rho$, similar to the proof of the deconfinement of $f$ anyons in Section \ref{sec:generalities}. The first criterion can be proved by contradiction.  We assume that $e^2m^2$ anyons can be locally created, i.e., $\exists V_{\partial{\tilde{C}}}$ supported near $\partial{\tilde{C}}$, s.t. $U_{\tilde{C}}\rho U^\dagger_{\tilde{C}}=V_{\partial\tilde{C}}\rho V^\dagger_{\partial\tilde{C}}$. Then we define two non-contractible Wilson loops $W^e_{\gamma_1}=\prod_{i\in\gamma_1}{Z_i},W^e_{\gamma_2}=\prod_{i\in\gamma_2}Z^\dagger_i$, as depicted in Fig.~\ref{fig:deconfinement}. Because $W^e_{\gamma_1}W^e_{\gamma_2}$ commute with $V_{\partial\tilde{C}}$ as well as all the stabilizers, we have
\begin{equation}
W^e_{\gamma_1}W^e_{\gamma_2}V_{\partial\tilde{C}}\rho V^\dagger_{\partial\tilde{C}}W^{e\dagger}_{\gamma_1}W^{e\dagger}_{\gamma_2}=V_{\partial\tilde{C}}\rho V^\dagger_{\partial\tilde{C}}.
\label{eq:local}
\end{equation}
On the other hand, $W^e_{\gamma_1}W^e_{\gamma_2}$ anticommutes with $W^{e^2m^2}_{\tilde{C}}$, so 
\begin{equation}
W^e_{\gamma_1}W^e_{\gamma_2}U_{\tilde{C}}\rho U^\dagger_{\tilde{C}}W^{e\dagger}_{\gamma_1}W^{e\dagger}_{\gamma_2}=U^\dagger_{\partial\tilde{C}}\rho U_{\partial\tilde{C}}\neq U_{\tilde{C}}\rho U^\dagger_{\tilde{C}},
\end{equation}
which leads to contradiction with \eqref{eq:local}. Therefore, $e^2m^2$ cannot be locally created, so they are deconfined excitations in the decohered double semion model.
\begin{figure}[htb]
 \centering
\includegraphics[width=0.6\linewidth]{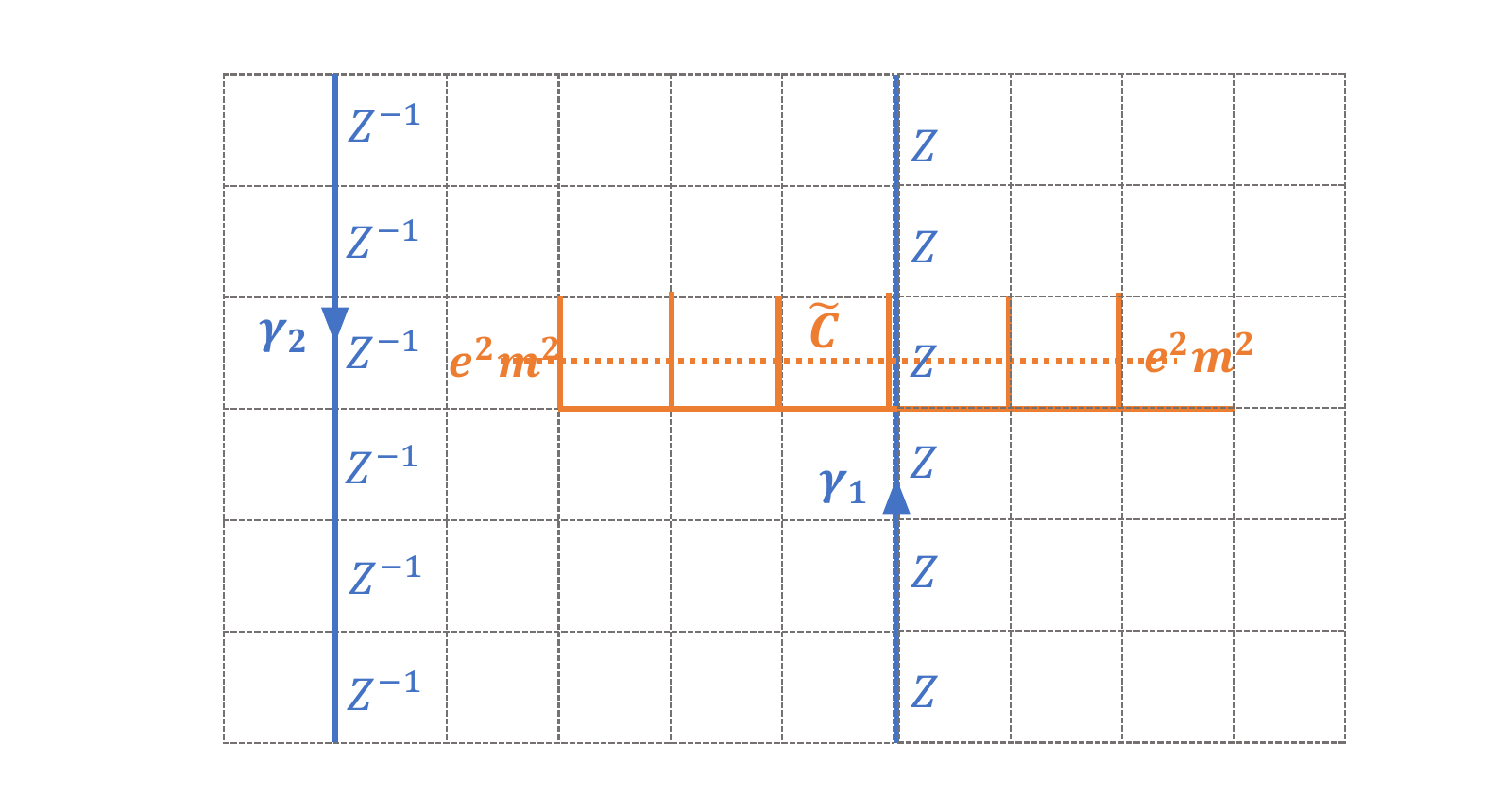}

\caption{Detecting $e^2m^2$ anyons using Wilson loops.}

\label{fig:deconfinement}
\end{figure}
\end{widetext}
\bibliography{wzj}
\end{document}